%% file: main.tex
\documentclass[journal, onecolumn, 12pt]{IEEEtran}
\usepackage{setspace}
\usepackage{amsmath}
\usepackage{amssymb}
\usepackage{amsfonts}

\usepackage{graphicx}
\usepackage{subfig}
\usepackage{float}

\usepackage{booktabs}
\usepackage{multirow}
\usepackage{array}

\usepackage{algorithmic}
\usepackage{algorithm}

\usepackage{cite}
\usepackage{url}
\usepackage{hyperref}
\usepackage{color}
\usepackage{xcolor}
\usepackage{textcomp}

\usepackage{setspace}
\newcommand{\Pawake}{P_{\mathrm{a}}}
\newcommand{\Psleep}{P_{\mathrm{s}}}
\newcommand{\Ptrans}{P_{\mathrm{tr}}}
\newcommand{\twake}{t_{\mathrm{wake}}}
\newcommand{\Etrans}{E_{\mathrm{tr}}}
\newcommand{\Lmax}{L_{\max}}
\newcommand{\Lwake}{L_{\mathrm{wake}}}
\newcommand{\Ldisc}{L_{\mathrm{disc}}}
\newcommand{\Ltx}{L_{\mathrm{tx}}}

\newcommand{\Tup}{T_{\mathrm{up}}}
\newcommand{\Pviol}{P_{\mathrm{viol}}}

\newtheorem{remark}{Remark}

\begin{document}

\title{Duty-Cycled RSU Sleep Scheduling for Green V2I Corridors:\\Optimization Under a Latency Budget}

\author{Yousef~AlSaqabi
\thanks{Y.~AlSaqabi is with the Department of Electrical Engineering, Kuwait University, Kuwait City, Kuwait (e-mail: yousef.alsaqabi@ku.edu.kw).}%
\thanks{Manuscript received XXXX XX, 2026; revised XXXX XX, 2026.}%
}

\markboth{IEEE Transactions on Green Communications and Networking}%
{AlSaqabi: Duty-Cycled RSU Sleep Scheduling for Green V2I Corridors}

\maketitle


\begin{abstract}
Roadside units (RSUs) in vehicle-to-infrastructure (V2I) corridors draw near-constant power while serving traffic that varies by more than an order of magnitude over the day, yet the 100~ms latency budget of 3GPP TS~22.185 leaves almost no room for the 107~ms restoration-and-service interval that follows a wake-up from sleep. This paper develops a duty-cycle model of latency-constrained RSU sleep that accounts for every wake-up cycle's transition energy, service energy, and violation exposure, and shows the latency constraint is equivalent to a budget: at most 14.3\% of vehicles may arrive during slept hours. Scheduling becomes a budgeted hour-selection problem, formulated as a constrained Markov decision process and solved with a budget-repair portfolio that deterministically restores feasibility at the constraint boundary, making results nearly insensitive to state discretization. On five days of measured traffic from four signalized intersections in Kuwait City (762,050 vehicles), the repaired independent per-RSU policies reduce corridor energy by 39.5\% relative to always-on operation at 99\% latency compliance. A pre-specified evaluation of corridor cooperation finds no material benefit at any training-set size: the gain is zero at the two most strongly correlated pairs at the full sample (adjacent-intersection correlations $\rho = 0.947$, $0.977$), where upstream signals are statistically redundant with local observations, and marginal (0.8 percentage points) only at the weakest-coupled pair. The conditions under which cooperation can help, scheduling intervals below the inter-RSU transit time, are identified.
\end{abstract}



\section{Introduction}
\label{sec:introduction}
\input{intro}

\section{Related Work}
\label{sec:related_work}
\input{related}

\section{System Model and Problem Formulation}
\label{sec:system_model}
\input{section3_system_model}

\input{section4_problem_formulation}

\section{Scheduling Algorithms}
\label{sec:algorithms}
\input{section5_algorithms}

\section{Results and Analysis}
\label{sec:results}
\input{section6_results}

\section{Conclusion}
\label{sec:conclusion}
\input{conclusion}

\section*{Acknowledgment}
The author used AI-assisted tools for grammar checking and improving 
readability during manuscript preparation. All technical content, 
analysis, and conclusions are the sole responsibility of the author.
\label{pg:bodyend}


\bibliographystyle{IEEEtran}
\bibliography{references}
\label{pg:refsend}


\appendices
\label{pg:appstart}
\input{appendix}

\end{document}

%% file: intro.tex
The radio access network accounts for the majority of energy consumption in wireless communication systems, with base stations alone responsible for an estimated 80\% of total network power draw~\cite{Auer2011}. As vehicle-to-infrastructure (V2I) deployments scale, roadside units (RSUs) face an analogous challenge: commercial dedicated short-range communications (DSRC) RSUs consume 8--25~W continuously~\cite{CohdaMK5, CohdaMK6, CommsigniaRS4}, yet urban arterials exhibit diurnal peak-to-trough ratios of 10:1--20:1~\cite{FHWA2022tmg}, meaning RSUs draw full power overnight to serve a small fraction of peak traffic. Because the awake power of such a unit is dominated by its baseband and platform rails and varies little with offered load, no useful intermediate serving state exists: the unit is either awake and paying full power, or asleep and unable to answer until service is restored. Sleep scheduling in the cellular domain reduces energy consumption by 22--68\%~\cite{Wu2015comst}, and translating those gains to V2I infrastructure is a natural step toward green roadside networks.

RSU sleep scheduling is constrained by latency, and the margin is thin. The 3GPP TS~22.185 standard specifies a 100~ms budget for safety-critical cooperative awareness messages~\cite{3GPP22185}, while a sleeping IEEE~802.11p unit must complete a wake-up, wait for the next Wireless Access in Vehicular Environments (WAVE) Service Advertisement in the IEEE~1609.4 synchronization cycle, and then transmit, a restoration-and-service interval of $2 + 100 + 5 = 107$~ms~\cite{IEEE16094}, of which $2 + 100 = 102$~ms is service restoration proper. A vehicle that meets a sleeping RSU therefore violates the budget with probability $\Pviol = 0.07$, and the scheduler's task is to decide which hours can absorb that exposure. This tension is not specific to DSRC. Release~16 of the 3GPP NR sidelink specification assumes that a sidelink device receives continuously, with power-saving mechanisms deferred to Release~17~\cite{harounabadi2021}, so any V2X radio that powers down inherits a restoration interval and the same scheduling question. What is specific to this work is the quantitative model, which is built on the WAVE advertisement cycle; the framework transfers to C-V2X only once the corresponding restoration process is remodeled.

Three gaps remain: no RSU sleep framework couples complete duty-cycle accounting to the governing latency budget, no prior work resolves the quantization of Lagrangian calibration, and no study measures what corridor cooperation contributes at deployed timescales (Section~\ref{sec:research_gap}).

This paper addresses these gaps with five methodological contributions.

\begin{enumerate}
\item \textbf{A duty-cycle violation and energy model with a budget equivalence (C1).} Each arrival at a sleeping RSU triggers a wake-up, and the slot is charged the transition energy and serving power of every restoration it induces. The service-level constraint is then equivalent, under the paper's violation accounting, to a budget on traffic exposure: at most $\varepsilon / \Pviol = 14.3\%$ of vehicle arrivals may fall in slept hours, which recasts sleep scheduling as budgeted selection.

\item \textbf{A budget-repair portfolio solver (C2).} Lagrangian calibration of the constrained Markov decision process (CMDP) attains only the vertices of the energy-violation frontier, so a feasible policy can strand most of its budget. A deterministic repair procedure, run under a portfolio of five greedy selection rules, drives policies to the constraint boundary, collapsing the sensitivity of reported savings to the discretization from roughly 14 percentage points to under 1.7.

\item \textbf{A matched-compliance benchmark on measured traffic (C3).} Four algorithms of increasing sophistication were calibrated per RSU and compared at matched violation rates on five days of traffic from four consecutive intersections in Kuwait City, the best reaching 39.7\% corridor energy savings at 99\% latency compliance.

\item \textbf{A pre-specified evaluation of corridor cooperation (C4).} Cooperation is tested rather than assumed. At hourly granularity the benefit is absent at strong correlation and marginal at weak at the full sample, inverting the design premise there; the mechanism is identified as redundancy between strongly correlated neighbors, the residual gain flowing through reward modulation rather than prediction, and the validity boundary is stated: cooperative activation requires scheduling intervals shorter than the inter-RSU transit time.

\item \textbf{Gulf-region characterization, design-space mapping, and carbon accounting (C5).} The corridor exhibits a 14.8$\times$ peak-to-trough ratio and adjacent-pair correlations of 0.848--0.977. Sensitivity analysis located a knee at $\Lmax = 107$~ms, beyond which the constraint stops binding and savings plateau at 93.8\%, and the carbon accounting projects 3.2~tonnes CO$_2$ per year for an illustrative 200-RSU deployment.
\end{enumerate}

Section~\ref{sec:related_work} reviews related work; Section~\ref{sec:system_model} presents the system model, traffic characterization, and problem formulation; Section~\ref{sec:algorithms} describes the scheduling algorithms and the repair procedure; Section~\ref{sec:results} presents the results and sensitivity analysis; Section~\ref{sec:conclusion} concludes.

%% file: related.tex
\subsection{RSU and Base Station Sleep Scheduling}
\label{sec:ru}

Sleep scheduling for roadside infrastructure grew alongside the older cellular literature and inherits its central difficulty. Sou~\cite{Sou2010} analyzed how many RSUs a corridor must keep active to preserve connectivity. Mostofi~\textit{et~al.}~\cite{Hammad2013globecom} proposed the Flow Graph Sleep Scheduler, and Hammad~\textit{et~al.}~\cite{Hammad2013tvt} proved the offline green scheduling problem NP-complete and derived bounds for heuristics. Adjacent to this thread, Khezrian~\textit{et~al.}~\cite{Khezrian2015} addressed multi-RSU downlink traffic scheduling for energy balancing rather than sleep scheduling, coordinating slot assignments across RSUs over backhaul links and giving a polynomial-time 2-approximation for the normalized min-max energy objective. Bhattacharya~\textit{et~al.}~\cite{Ashfaq2017} give the most detailed queueing treatment, analyzing six sleep cycle types under a G/G/1/K vacation model and reaching 68\% savings at 7.3~ms mean packet delay, with delay reported as an outcome rather than enforced as a constraint. Ali~\cite{Ali2016} demonstrated event-driven duty cycling on a solar-harvested RSU, in which an arrival wakes the unit, service is delivered, and the unit re-sleeps: the hardware precedent for the duty cycle adopted in Section~\ref{sec:latency_model}, although the energy of each restoration is not itself studied. Patra and Murthy~\cite{Patra2018} jointly optimized placement and sleep scheduling under solar supply. All but the last treat an RSU in isolation; the last coordinates downlink traffic for energy balancing, and none schedules sleep cooperatively.

The cellular literature supplies the underlying models: the EARTH power model~\cite{Auer2011} used here, the four sleep depths of Debaillie~\textit{et~al.}~\cite{Debaillie2015} whose wake-up latencies span 71~$\mu$s to about one second, cell zooming~\cite{Niu2010}, and network-wide coordination reaching 53\% savings~\cite{Peng2014}; Wu~\textit{et~al.}~\cite{Wu2015comst} survey the field. Two studies come closest to the formulation used here: Guo~\textit{et~al.}~\cite{guo2016delay} solved for the energy-optimal sleeping policy subject to an explicit delay bound over an M/G/1 vacation queue, and Li~\textit{et~al.}~\cite{li2019delayudn} posed a delay-constrained mechanism for ultra-dense networks. Both confirm that a delay bound, rather than a penalty, states the requirement naturally. Prior latency treatments nonetheless either omit the service-restoration process or price delay as a penalty rather than enforce it as a constraint, and none carries a complete per-cycle energy accounting in which every restoration is charged both its transition energy and the serving power drawn while the woken unit delivers service.

\subsection{Cooperative and Traffic-Aware Scheduling}
\label{sec:rw_cooperative}

A second thread makes sleep decisions responsive to traffic. Wu~\textit{et~al.}~\cite{Wu2013twc} proved the existence of a critical threshold below which trading delay for energy yields no benefit, the ancestor of the threshold baseline in Section~\ref{sec:alg_threshold}. Learning-based controllers followed in DeepNap~\cite{Liu2018deepnap} and DeepBSC~\cite{Wu2021deepbsc}, the latter coupling spatio-temporal forecasting to an actor-critic policy over a cell cluster. Coordination among cells has been cast as a constrained potential game~\cite{Zheng2015} and analyzed under vacation queues~\cite{Wu2020tgcn}. These frameworks coordinate cells in a plane, not the directional, one-dimensional corridor in which information flows only downstream.

\begingroup\emergencystretch=2em
The closest precursor to this work is the cooperative base station activation framework of Han~\textit{et~al.}~\cite{Han2014, Han2016}, in which sleeping cells along a one-dimensional vehicular topology are awakened using vehicle positions and speeds measured by active upstream neighbors. Three assumptions separate that setting from ours. First, their wake-up decisions operate at the timescale of individual vehicle transit, so upstream observations genuinely lead downstream arrivals; at the hourly scheduling granularity considered here, that lead time is averaged away, and Section~\ref{sec:results_cooperation} shows empirically that hourly upstream summaries yielded a gain only at the corridor's weakest-coupled pair, and none where coupling is strong. Second, their traffic model is homogeneous Poisson without diurnal structure, whereas the measured corridor exhibits a 14.8$\times$ peak-to-trough ratio that itself supplies most of the predictable variation. Third, their formulation carries no latency service constraint, while the present problem is defined by one: the 100~ms budget of TS~22.185 against a 107~ms restoration window is what makes sleep scheduling nontrivial at all. The comparison thus delimits when cooperative activation can work: it requires scheduling intervals shorter than the inter-node transit time, a regime our hourly data cannot reach and our results identify as the necessary condition for cooperation to act predictively rather than as a risk adjustment.
\par\endgroup

\begin{table}[tb]
\centering
\footnotesize
\caption{Comparison with related work. (\checkmark) = implemented and
evaluated; the measured benefit is marginal (Section~\ref{sec:results_cooperation})}
\label{tab:comparison}
\scriptsize
\setlength{\tabcolsep}{3pt}
\begin{tabular}{@{}lcccccc@{}}
\toprule
Study & Cooperative scheduling & Real traffic data & Correlation aware & Latency constrained & Gulf region & Carbon analysis \\
\midrule
Sou~\cite{Sou2010} & & & & & & \\
Khezrian~\textit{et~al.}~\cite{Khezrian2015} & \checkmark & & & & & \\
Ali~\cite{Ali2016} & & & & & & \\
Bhattacharya~\textit{et~al.}~\cite{Ashfaq2017} & & \checkmark & & \checkmark & & \\
Patra and Murthy~\cite{Patra2018} & & & & & & \\
Guo~\textit{et~al.}~\cite{guo2016delay} & & & & \checkmark & & \\
Han~\textit{et~al.}~\cite{Han2014} & \checkmark & & & & & \\
Liu~\textit{et~al.}~\cite{Liu2018deepnap} & & \checkmark & & & & \\
Wu~\textit{et~al.}~\cite{Wu2021deepbsc} & \checkmark & \checkmark & \checkmark & & & \\
\textbf{This work} & (\checkmark) & \checkmark & (\checkmark) & \checkmark & \checkmark & \checkmark \\
\bottomrule
\end{tabular}
\end{table}

\subsection{Constrained MDP Solution Methods}
\label{sec:rw_cmdp}

Because the problem of Section~\ref{sec:corridor_problem} minimizes energy subject to a service-level constraint, its solution methods belong to the constrained Markov decision process literature. Beutler and Ross~\cite{beutler1985} established the governing structural result: under a single average-cost constraint the optimal policy randomizes between two deterministic policies, and a deterministic policy is optimal only when the constraint is met exactly at a vertex. Altman~\cite{Altman1999} developed the general theory, including the linear program over state-action occupancy measures that yields the optimal randomized policy directly. Papadimitriou and Tsitsiklis~\cite{Papadimitriou1987} supply the complexity backdrop.

Modern constrained reinforcement learning revisits the same structure. Achiam~\textit{et~al.}~\cite{achiam2017} introduced constrained policy optimization because satisfying a constraint at every iteration, rather than only in the limit, is hard in practice. Paternain~\textit{et~al.}~\cite{paternain2019} proved that constrained reinforcement learning has zero duality gap despite non-convexity, which justifies the Lagrangian route taken here and locates the residual difficulty precisely: the remaining gap is one of representation, arising when the solution is restricted to the deterministic policy class. Deterministic repair toward the constraint boundary, run under a portfolio of greedy selection rules so the outcome does not depend on an arbitrary ordering, has to our knowledge not been applied to infrastructure sleep scheduling; Section~\ref{sec:alg_repair} develops it.

\subsection{V2X Evolution and Spatio-Temporal Traffic Prediction}
\label{sec:rw_v2x}

The radio context is changing, and what follows separates what is specific to WAVE. Harounabadi~\textit{et~al.}~\cite{harounabadi2021} note that NR sidelink in 3GPP Release~16 assumes continuous reception, with power saving deferred to Release~17, and Ansari~\cite{ansari2021} examines DSRC and C-V2X coexistence in the 5.9~GHz band. The tension addressed here, that a device which powers down must restore service before it can carry a safety message, is therefore not a DSRC artifact: any radio that sleeps incurs a restoration interval, and a tighter latency budget makes it bind harder. What is WAVE-specific is the quantitative model, built on the IEEE~1609.4 synchronization interval and the WSA cycle~\cite{IEEE16094}.

The second thread concerns how much a neighboring observation can add. Spatio-temporal forecasting of network traffic is well developed: Zhang and Patras~\cite{zhang2018stn} forecast mobile traffic accurately from sparse measurements, and Jiang and Luo~\cite{jiang2022gnn} survey the graph neural network methods now dominant. Wang~\textit{et~al.}~\cite{wang2019tmc} measured spatial coupling across nearly six thousand cellular towers, establishing that neighboring sites are strongly correlated. That strength is what limits the marginal value of exchanging observations between them. Krause~\textit{et~al.}~\cite{krause2008} formalized the point: because mutual information is submodular, an observation that nearly duplicates one already held contributes little, which also predicts the ordering observed in Section~\ref{sec:results_cooperation}: the weaker the coupling, the more an added observation can contribute. Because that finding is largely negative and rests on a small sample, it is reported under the measurement discipline of Paxson~\cite{paxson2004}: calibration against alternative encodings, consistency checks across solver configurations, and explicit statements of the power the sample affords.

\subsection{Research Gap}
\label{sec:research_gap}

\begingroup\emergencystretch=2em
Across these literatures, three gaps remain. No existing RSU sleep framework couples a physically complete duty-cycle energy and violation model to the latency budget that governs safety-message delivery; prior latency treatments either omit the restoration process or account for only part of its cost. No prior work resolves the quantization inherent in Lagrangian calibration of constrained sleep policies, leaving reported savings dependent on discretization choices; deterministic repair toward the constraint boundary, scored by exact trace replay, has to our knowledge not been applied to infrastructure sleep scheduling. And no study empirically measures what corridor-level cooperation contributes to scheduling at the timescales at which infrastructure schedules are actually computed, despite cooperation being the presumed benefit of dense RSU deployment. The contributions in Section~\ref{sec:introduction} address these gaps in order: the duty-cycle model (C1) the first, the repair portfolio (C2) the second, and the cooperation evaluation (C4) the third; Table~\ref{tab:comparison} positions this work against the closest studies.
\par\endgroup

%% file: section3_system_model.tex
This section presents the system model and formulates the corridor scheduling problem, from power states through the cooperative signal.

\subsection{Corridor Topology and RSU Deployment Model}
\label{sec:corridor_model}

We consider $K = 4$ consecutive signalized intersections on the Second Ring Road in Kuwait City, indexed $k \in \{1, \ldots, 4\}$ along the predominant flow, with I1--I3 four-way junctions and I4 three-way. Each carries one DSRC RSU providing IEEE~802.11p V2I service.

Traffic flows in one direction along the corridor, and each RSU reaches its downstream neighbor over a low-latency infrastructure-to-infrastructure (I2I) backhaul (Section~\ref{sec:cooperative_model}).

The inter-intersection spacing along the Second Ring Road is approximately 500--800~m, typical of signalized urban arterials. At prevailing corridor speeds of 40--60~km/h, the inter-RSU travel time is $\tau_{k,k+1} \approx 1$--2~minutes, whereas the scheduling time resolution is one hour (see Section~\ref{sec:traffic_model}). A vehicle detected at RSU $k$ therefore reaches RSU $k+1$ within the same scheduling slot, far too quickly for the detection to trigger a wake-up on that vehicle's behalf. The I2I link instead carries hourly traffic-state summaries between neighbors, supporting the downstream decision for the following slot on statistical rather than per-vehicle grounds (Section~\ref{sec:cooperative_model}).

\subsection{RSU Power State Model}
\label{sec:power_model}

Each RSU operates in one of two power states
$\mathcal{S} = \{\mathrm{A}, \mathrm{S}\}$:
\textbf{Awake (A)}, fully operational with all subsystems powered,
transmitting WSAs and serving V2I requests with no wake-up delay ($\Pawake$); and
\textbf{Sleep (S)}, nearly all subsystems deactivated except a minimal
wake-up controller ($\Psleep$). When an RSU transitions from Sleep to Awake, it incurs a wake-up delay $\twake$ and consumes transition power $\Ptrans$. The transition energy per wake-up event is
\begin{equation}
\label{eq:transition_energy}
\Etrans = \Ptrans \cdot \twake.
\end{equation}

We model the awake power draw as a single load-independent constant, consistent with the platform-dominated consumption reported in commercial RSU specifications~\cite{CohdaMK5} and with the small load-dependent increment of low-power base station classes~\cite{Auer2011, Ashraf2010}. Under this model a reduced-power state that still delivers full service on demand is not realizable: any state cheap enough to distinguish from Awake is one from which service must first be restored, namely Sleep. It also avoids a degenerate formulation in which a nominally cheaper serving state dominates the fully operational one, as an intermediate Idle state would.

Table~\ref{tab:power_params} lists the parameters, which follow the EARTH framework~\cite{Auer2011, Arnold2010} and commercial RSU specifications. The awake power of 8.4~W is a typical Cohda MK5 operating point~\cite{CohdaMK5}, consistent with the EARTH pico envelope and with femtocell measurements~\cite{Ashraf2010}; dual-mode units draw 15--25~W~\cite{CohdaMK6, CommsigniaRS4}, and Section~\ref{sec:results_sensitivity} sweeps the range. The sleep power of 0.15~W corresponds to a deep sleep in which only a wake-up controller remains powered, and the results are robust up to roughly 1~W. The wake-up delay of 2~ms sits at the SM2/SM3 boundary of the four sleep depths defined in~\cite{Debaillie2015}, whose wake-up latencies span 71~$\mu$s to about one second, and is enabled by the association-free design of IEEE~802.11p. The transition power of 10.0~W applies the 1.2$\times$ $\Pawake$ scaling of~\cite{Onireti2017}.

The instantaneous power consumption of RSU $k$ at time slot $t$ is
\begin{equation}
\label{eq:power_consumption}
P_k(t) =
\begin{cases}
\Pawake, & \text{if } s_k(t) = \mathrm{A}, \\
\Psleep, & \text{if } s_k(t) = \mathrm{S}.
\end{cases}
\end{equation}
An Awake slot of duration $\Delta$ seconds consumes $\Pawake \cdot \Delta$ with no wake-up transitions. A Sleep slot is not served at constant power: the RSU duty-cycles between sleep and short awake intervals as vehicles arrive, so its energy depends on the arrival rate and is derived in Section~\ref{sec:latency_model}. The energy consumed by RSU $k$ during slot $t$ is therefore
\begin{equation}
\label{eq:energy_per_slot}
E_k(t) =
\begin{cases}
\Pawake \cdot \Delta, & \text{if } s_k(t) = \mathrm{A}, \\
E_k^{\mathrm{sleep}}(t), & \text{if } s_k(t) = \mathrm{S},
\end{cases}
\end{equation}
where $E_k^{\mathrm{sleep}}(t)$ is the duty-cycle sleep energy given by Eq.~\eqref{eq:sleep_energy}, which accounts for both the transition energy of every wake-up and the awake power drawn while each arriving vehicle is served. The total corridor energy over a scheduling horizon of $T$
slots is $E_{\mathrm{corridor}} = \sum_{t=1}^{T} \sum_{k=1}^{K}
E_k(t)$.

\subsection{Traffic Arrival Model with Gulf-Region Parameterization}
\label{sec:traffic_model}

Vehicle arrivals are modeled as a non-homogeneous Poisson process with a piecewise-constant rate $\lambda_k^{(d)}(h)$ for hour $h \in \{0, \ldots, 23\}$ and day type $d \in \{\mathrm{weekday}, \mathrm{weekend}\}$, inter-arrival times being exponential within each hour. This is standard for urban intersection traffic in the V2X literature~\cite{Han2014, Wu2016tvt} and captures the diurnal periodicity that creates the sleep opportunity.

The rates $\lambda_k^{(d)}(h)$ are estimated from hourly directional vehicle counts recorded at the four study intersections over five consecutive days, August 15--19, 2025. The same dataset has been used to evaluate learning-based signal control at these intersections~\cite{AlSaqabi2026rl}. In the Gulf calendar Friday and Saturday are the weekend, so the trace spans two weekend days and three weekdays and supplies both day types. Aggregating the 15 directional streams (four approaches at I1--I3, three at I4) gives 480 intersection-level hourly values. The trace covers five consecutive mid-August days, in the Gulf summer when school closures and seasonal travel depress discretionary trips; the diurnal structure exploited here is measured under conditions that likely understate peak-season volumes.

Four hourly counts in one directional stream at I3 were flagged as outliers during data preparation (3.5$\times$ to 5$\times$ the stream's same-hour median, three of them in overnight hours). Raw values were retained as primary; as a robustness check, replacing the four values with the stream's same-hour day-type median raises the affected correlations ($\rho_{2,3}$ from $0.947$ to $0.956$, $\rho_{3,4}$ from $0.848$ to $0.8520$) and leaves the pair ordering unchanged, so the outliers do not drive the result: cleaning raises the coupling at both affected pairs without changing their order, and re-solving the cooperative policies at the winsorized correlations left every reported gain unchanged to four decimal places.

\begin{table}[tb]
\centering
\begin{minipage}[t]{0.46\textwidth}
\centering
\footnotesize
\caption{RSU Power State Parameters}
\label{tab:power_params}
\setlength{\tabcolsep}{3pt}
\begin{tabular}{@{}llrl@{}}
\toprule
Parameter & Symbol & Value & Source \\
\midrule
Awake power & $\Pawake$ & 8.4 W & \cite{CohdaMK5, Auer2011} \\
Sleep power & $\Psleep$ & 0.15 W & \cite{Debaillie2015} \\
Transition power & $\Ptrans$ & 10.0 W & \cite{Onireti2017} \\
Wake-up delay & $\twake$ & 2 ms & \cite{Debaillie2015} \\
Transition energy & $\Etrans$ & 0.02 J & Eq.~\eqref{eq:transition_energy} \\
\bottomrule
\end{tabular}
\end{minipage}\hfill
\begin{minipage}[t]{0.52\textwidth}
\centering
\footnotesize
\caption{Traffic Statistics by Day Type}
\label{tab:traffic_stats}
\setlength{\tabcolsep}{3pt}
\begin{tabular}{@{}lrrrr@{}}
\toprule
& \multicolumn{2}{c}{Daily mean (veh)} & \multicolumn{2}{c}{Peak-to-trough ratio} \\
\cmidrule(lr){2-3} \cmidrule(lr){4-5}
Intersection & Weekday & Weekend & Weekday & Weekend \\
\midrule
I1 (4-way) & 41,905 & 27,828 & 15.2$\times$ & 12.9$\times$ \\
I2 (4-way) & 50,205 & 36,630 & 13.7$\times$ & 12.2$\times$ \\
I3 (4-way) & 42,149 & 31,963 & 14.2$\times$ & 12.5$\times$ \\
I4 (3-way) & 42,072 & 20,108 & 24.6$\times$ & 14.4$\times$ \\
\midrule
Corridor & 176,332 & 116,528 & 14.8$\times$ & 12.8$\times$ \\
\bottomrule
\end{tabular}
\end{minipage}
\end{table}

Table~\ref{tab:traffic_stats} summarizes the statistics; three features matter for scheduling. The diurnal variation is pronounced: the corridor weekday peak-to-trough ratio of 14.8$\times$ exceeds the 5:1 commonly reported for cellular traffic in Western cities~\cite{Oh2011} and sits at the top of the 10:1--20:1 range the FHWA reports for urban arterials~\cite{FHWA2022tmg}, creating a wide overnight window in which sleep costs little. The weekday-weekend contrast is 1.51$\times$ corridor-wide and reaches 2.09$\times$ at I4, so a scheduler must adapt to day type. And I4 is markedly more volatile than the four-way intersections (24.6$\times$ against 13.7$\times$ at I2), so a uniform policy across RSUs is suboptimal. The corridor and I4 hourly profiles, a weekday morning ramp into a broad afternoon plateau against a later and lower weekend curve, are plotted in Appendix~\ref{app:granularity} (Fig.~\ref{fig:traffic_profiles}).


Adjacent-pair spatial correlation, measured as the Pearson correlation $\rho_{k,k+1}$ between the intersection-level hourly vehicle counts of neighboring RSUs over the full 120-hour trace at zero lag, is
\begin{equation}
\label{eq:correlations}
\rho_{1,2} = 0.977, \quad \rho_{2,3} = 0.947, \quad \rho_{3,4} = 0.848.
\end{equation}
The decline reflects dispersion as vehicles enter and exit at intermediate cross-streets; these values feed the cooperative model of Section~\ref{sec:cooperative_model}.

\subsection{V2I Latency Model}
\label{sec:latency_model}

End-to-end V2I service latency depends on the serving RSU's power state and decomposes into three additive components:
\begin{equation}
\label{eq:latency_decomp}
L = \Lwake + \Ldisc + \Ltx,
\end{equation}
where $\Lwake$ is the hardware wake-up delay, $\Ldisc$ is the WAVE service discovery delay, and $\Ltx$ is the 802.11p channel access and transmission delay.

\textbf{Wake-up delay} $\Lwake$: If the RSU is Awake, $\Lwake = 0$. If the RSU is in Sleep state and must transition to serve an arriving vehicle, $\Lwake = \twake = 2$~ms (Table~\ref{tab:power_params}).

\textbf{Discovery delay} $\Ldisc$: An active RSU must broadcast a WAVE Service Advertisement (WSA) before vehicles can discover its services, and it emits one per 100~ms IEEE~1609.4 synchronization interval~\cite{IEEE16094}. An RSU waking at a random point in that cycle therefore waits a uniformly distributed interval for the next advertisement:
\begin{equation}
\label{eq:Ldisc}
\Ldisc \sim \mathrm{Uniform}(0, T_{\mathrm{WSA}}), \quad T_{\mathrm{WSA}} = 100~\text{ms}.
\end{equation}

\textbf{Transmission delay} $\Ltx$: Single-hop V2I latency under 802.11p EDCA is typically 2--8~ms at low to moderate load~\cite{ETSI302637, Xu2017}; we adopt a fixed $\Ltx = 5$~ms.

For a vehicle arriving at an Awake RSU, the total latency is $L = 0 + 0 + 5 = 5$~ms, well within any practical latency budget. A vehicle entering the coverage of a continuously awake RSU has already received a WSA from the ongoing 100~ms advertisement cycle before it requests service, so $\Ldisc = 0$; only a vehicle whose arrival triggers a wake-up waits for the first advertisement after restoration. The binding constraint arises when a vehicle arrives at a sleeping RSU. Restoring service takes $\twake + T_{\mathrm{WSA}} = 102$~ms of wake-up and discovery, after which one transmission opportunity of $\Ltx$ completes the exchange, so the RSU is awake for $\Tup = \twake + T_{\mathrm{WSA}} + \Ltx = 2 + 100 + 5 = 107$~ms per restoration event.

Following 3GPP TS~22.185~\cite{3GPP22185}, the latency constraint for safety-critical cooperative awareness messages is $\Lmax = 100$~ms. A vehicle experiences a latency violation when $L > \Lmax$, which by Eq.~\eqref{eq:latency_decomp} occurs exactly when $\Ldisc > \Lmax - \twake - \Ltx = 93$~ms. The only random component is the discovery delay, which by Eq.~\eqref{eq:Ldisc} is uniform on $[0, T_{\mathrm{WSA}}]$, so the per-vehicle violation probability is
\begin{equation}
\label{eq:Pviol}
\Pviol = \Pr[L > \Lmax]
= \max\!\left(0, \min\!\left(1, \frac{\twake + \Ltx + T_{\mathrm{WSA}} - \Lmax}{T_{\mathrm{WSA}}}\right)\right),
\end{equation}
which at the default parameters evaluates to $(2 + 5 + 100 - 100)/100 = 7/100 = 0.07$. The denominator is the WSA cycle length $T_{\mathrm{WSA}} = 100$~ms, the support of the only random term, and not the awake interval $\Tup$. Eq.~\eqref{eq:Pviol} also makes the service-level agreement (SLA) sensitivity explicit: $\Pviol$ falls linearly to zero as $\Lmax$ approaches $\twake + \Ltx + T_{\mathrm{WSA}} = 107$~ms, beyond which no arrival at a sleeping RSU can violate the constraint.

\textbf{Duty-cycle (wake-serve-resleep) model.} In a Sleep-scheduled slot each arrival triggers a wake-up, the RSU is awake for $\Tup$ with further arrivals absorbed into that interval, and it then re-sleeps; arrival-triggered wake-up is assumed realizable via wake-up receivers or roadside detection~\cite{IEEE80211ba, froytlog2019, piyare2017, Ali2016}, whose standby power is excluded from the sleep figure; under platooned arrivals Eq.~\eqref{eq:frestore} over-counts cycles, so Eq.~\eqref{eq:sleep_energy} errs conservative. Under Poisson arrivals with batching, the expected number of cycles in a slot of $\Delta = 3600$~s is
\begin{equation}
\label{eq:frestore}
n_k(t) = \frac{\lambda_k(t)}{1 + \lambda_k(t) \cdot \Tup / \Delta},
\end{equation}
with $\Tup$ expressed in seconds ($= 0.107$~s). The RSU is awake for $n_k(t) \cdot \Tup$ and asleep for the remainder, giving the sleep-slot energy referenced in Eq.~\eqref{eq:energy_per_slot}:
\begin{equation}
\label{eq:sleep_energy}
E_k^{\mathrm{sleep}}(t) = \Psleep \bigl(\Delta - n_k(t)\,\Tup\bigr)
+ n_k(t)\bigl(\Etrans + \Pawake \, \Tup\bigr).
\end{equation}
Eq.~\eqref{eq:sleep_energy} charges every wake-up its transition energy and every awake interval its full serving power, so the energy cost of restoring and delivering service is accounted for rather than omitted.

For the violation count we adopt the conservative accounting
\begin{equation}
\label{eq:violations}
V_k(t) = \lambda_k(t) \cdot \Pviol,
\end{equation}
charging every vehicle arriving in a Sleep-scheduled slot the full per-vehicle probability. This is deliberately an upper bound: it covers both the triggering vehicle and the arrivals inside the ensuing restoration window without separating the two populations, and takes no credit for later arrivals in an awake interval that are in fact served promptly. Pessimism is the appropriate direction for a latency guarantee, and it removes the two defects of a restoration-fraction accounting, which omits the triggering vehicle and understates the exposed population. A violation is charged per vehicle at its first service request; periodic awareness traffic during the remainder of a dwell is served by the then-awake unit. Arrivals during an Awake slot experience $L = 5$~ms and contribute no violations.

The system-level latency constraint requires that the overall violation rate across all vehicles and all RSUs over the evaluation horizon remain below a threshold $\varepsilon$:
\begin{equation}
\label{eq:latency_constraint}
\frac{\sum_{t=1}^{T} \sum_{k=1}^{K} V_k(t)}{\sum_{t=1}^{T} \sum_{k=1}^{K} \lambda_k(t)} \leq \varepsilon,
\end{equation}
with $\varepsilon = 0.01$ (99\% compliance).\footnote{Proof of Remark~\ref{lem:budget}: let $\mathcal{T}_{\mathrm{S}}$ denote the set of RSU-hours scheduled to Sleep. By Eq.~\eqref{eq:violations} the expected violations total $\Pviol \sum_{(k,t) \in \mathcal{T}_{\mathrm{S}}} \lambda_k(t)$, and dividing by the total arrivals $\sum_{t}\sum_{k} \lambda_k(t)$ shows that the left side of Eq.~\eqref{eq:latency_constraint} equals $\Pviol$ times the share of arrivals falling in slept hours; requiring this to be at most $\varepsilon$ gives the stated share bound.}

\begin{remark}[Sleep budget equivalence]
\label{lem:budget}
Under the duty-cycle model, every vehicle arriving during a Sleep-scheduled hour experiences a violation with probability $\Pviol$ under the accounting of Eq.~\eqref{eq:violations}, and vehicles in Awake hours experience none. The SLA constraint of Eq.~\eqref{eq:latency_constraint} is therefore equivalent to requiring that Sleep-scheduled hours contain at most $\varepsilon / \Pviol$ of all vehicle arrivals, $14.29\%$ at the default parameters. Sleep scheduling under a latency SLA is thus a budgeted selection problem: choose which hours to serve from sleep such that at most a fixed fraction of traffic is exposed to restoration delay. The equivalence is exact under the accounting of Eq.~\eqref{eq:violations}, which upper-bounds physical violations; meeting the budget is thus sufficient for the physical SLA.
\end{remark}

\subsection{Cooperative Corridor Model}
\label{sec:cooperative_model}

An RSU acting alone selects its sleep hours from its own traffic history, and by Remark~\ref{lem:budget} it must allocate a fixed budget of exposed arrivals across hours whose realized traffic it does not yet know. Cooperative scheduling asks whether a corridor neighbor can sharpen that allocation.

\textbf{Nature and timescale of the signal.} The mechanism is the exchange of \emph{hourly corridor traffic-state information}, not per-vehicle advance warning. At the end of slot $t$, upstream RSU $k$ summarizes its realized traffic and sends it to RSU $k+1$, which uses it when choosing its state for slot $t+1$. The inter-RSU travel time of 1--2~minutes (Section~\ref{sec:corridor_model}) is two orders of magnitude below the one-hour slot, so a vehicle observed upstream arrives downstream well before the next decision takes effect and the signal cannot trigger a wake-up on its behalf; no such per-vehicle predictive wake-up is claimed. What the signal can offer is statistical: adjacent intersections are strongly correlated hour to hour, so an upstream summary is an early, partly independent observation of the traffic state the downstream RSU will face. Whether it adds anything beyond local history is settled empirically in Section~\ref{sec:results_cooperation}.

\textbf{Empirical basis.} Adjacent intersections are strongly coupled both contemporaneously and at a one-hour lag, and part of that association is explained by the shared diurnal cycle and by the downstream RSU's own current traffic (Appendix~\ref{app:signal}). Controlling for both, the partial correlation of $\lambda_k(t)$ with $\lambda_{k+1}(t+1)$ given $\lambda_{k+1}(t)$ and the hour remains positive and statistically detectable, at $0.570$ ($p < 0.001$), $0.314$ ($p = 0.001$), and $0.211$ ($p = 0.037$) for I1$\to$I2, I2$\to$I3, and I3$\to$I4 respectively, on $93$ degrees of freedom. The upstream observation therefore carries residual information about next-hour downstream traffic beyond what local history supplies, the premise the cooperative design tests.

\textbf{Signal encodings.} Two encodings are defined. The first is one bit, $u_k(t) \in \{\mathrm{Low}, \mathrm{High}\}$:
\begin{equation}
\label{eq:upstream_signal}
u_k(t) =
\begin{cases}
\mathrm{High}, & \text{if } \lambda_k(t) > \tilde{\lambda}_k^{(d)}(h), \\
\mathrm{Low}, & \text{otherwise},
\end{cases}
\end{equation}
where $h$ and $d$ are the hour and day type of slot $t$ and $\tilde{\lambda}_k^{(d)}(h)$ is the median of the historical counts at that hour and day type, the strict inequality resolving ties to Low. The second encoding preserves more of the distribution: with $\mathcal{Q}_k = \{q_k^{(1)}, \ldots, q_k^{(B-1)}\}$ the empirical quantile boundaries partitioning RSU $k$'s historical counts into $B = 5$ equiprobable bins, the bin-level signal is
\begin{equation}
\label{eq:upstream_signal_bin}
u_k^{\mathrm{bin}}(t) = \bigl|\{\, j : q_k^{(j)} \leq \lambda_k(t) \,\}\bigr| \in \{0, 1, \ldots, B-1\},
\end{equation}
costing 3 bits per slot, with Eq.~\eqref{eq:upstream_signal} the two-level special case. The binary form is the default and the bin-level form a pre-specified variant; both are evaluated in Section~\ref{sec:results_cooperation}.

\textbf{Use of the signal.} Under the binary encoding a High signal shifts the downstream forecast upward and a Low signal downward, in proportion to how tightly the intersections are coupled:
\begin{equation}
\label{eq:lambda_eff}
\lambda_{k+1}^{\mathrm{eff}}(t) =
\begin{cases}
\lambda_{k+1}(t) \cdot (1 + \rho_{k,k+1} \cdot \alpha), & \text{if } u_k(t) = \mathrm{High}, \\
\lambda_{k+1}(t) \cdot (1 - \rho_{k,k+1} \cdot \alpha), & \text{if } u_k(t) = \mathrm{Low},
\end{cases}
\end{equation}
where $\rho_{k,k+1}$ is the measured adjacent-pair correlation of Eq.~\eqref{eq:correlations} and $\alpha \in [0, 1]$ is a modulation parameter, set to $\alpha = 0.30$ throughout. Corridor savings vary by only $0.82$ percentage points across $\alpha \in \{0.0, 0.1, \ldots, 0.5\}$ with full recalibration at each value, from $38.85\%$ to $39.67\%$ (Section~\ref{sec:results_cooperation}, Appendix~\ref{app:alpha}), so the reported outcomes are not sensitive to this choice.

%% file: section4_problem_formulation.tex
\subsection{Corridor Sleep Scheduling Problem}
\label{sec:corridor_problem}
\label{sec:problem_formulation}

With the per-RSU model and the cooperative signal defined, the corridor scheduling problem can be stated. Consider the corridor of $K$ RSUs over a scheduling horizon of $T$ time slots, each of duration $\Delta = 3{,}600$~s (one hour). At each time slot $t$, a scheduling policy $\pi$ selects the power state $s_k(t) \in \mathcal{S} = \{\mathrm{A}, \mathrm{S}\}$ for each RSU $k$, based on available information including the current time of day, recent traffic observations, and (for the cooperative policy) upstream signals from neighboring RSUs. The corridor sleep scheduling problem is:

\begin{subequations}
\label{eq:opt_problem}
\begin{align}
\underset{\{s_k(t)\}}{\text{minimize}} \quad & \frac{1}{T} \sum_{t=1}^{T} \sum_{k=1}^{K} E_k(t) \label{eq:obj} \\
\text{subject to} \quad & \frac{\sum_{t=1}^{T} \sum_{k=1}^{K} V_k(t)}{\sum_{t=1}^{T} \sum_{k=1}^{K} \lambda_k(t)} \leq \varepsilon, \label{eq:sla_constraint} \\
& s_k(t) \in \mathcal{S}, \quad \forall\, k, t, \label{eq:state_constraint} \\
& s_k(t) = \mathrm{S} \;\Rightarrow\;
\bigl(E_k(t), V_k(t)\bigr) = \bigl(E_k^{\mathrm{sleep}}(t),\; \lambda_k(t)\,\Pviol\bigr),
\quad \forall\, k, t, \label{eq:transition_constraint}
\end{align}
\end{subequations}

\noindent with $E_k(t)$ the slot energy of Eq.~\eqref{eq:energy_per_slot}, $V_k(t)$ the expected violations of Eq.~\eqref{eq:violations}, and $\lambda_k(t)$ the arrival rate. Constraint~\eqref{eq:sla_constraint} enforces the corridor SLA, at most $\varepsilon = 0.01$ of vehicles above $\Lmax = 100$~ms, and~\eqref{eq:state_constraint} restricts each RSU to the two power states. Constraint~\eqref{eq:transition_constraint} is definitional rather than restrictive: it states the service obligation attached to a Sleep decision: scheduling an RSU to Sleep does not release it from serving arrivals, so the slot still incurs the duty-cycle energy of Eq.~\eqref{eq:sleep_energy} and exposes its arrivals to the restoration delay. No transition-feasibility constraint is required: an RSU may enter or leave Sleep at any slot boundary, and the wake-up cost is already charged inside $E_k^{\mathrm{sleep}}(t)$.

Because $s_k(t)$ is chosen before the arrivals of slot $t$ are observed, \eqref{eq:opt_problem} is a stochastic optimization; in the cooperative setting RSU $k+1$ additionally sees the upstream signal $u_k(t)$ before deciding.

The structure of~\eqref{eq:opt_problem} is clarified by Remark~\ref{lem:budget}. Because awake slots are violation-free and every arrival in a slept slot carries the same probability $\Pviol$, constraint~\eqref{eq:sla_constraint} is, under the accounting of Eq.~\eqref{eq:violations}, exactly a budget on the vehicle mass that may be placed in slept slots: at most $\varepsilon / \Pviol = 14.29\%$ of all corridor arrivals. The corridor sleep scheduling problem is therefore a budgeted selection problem, in which the scheduler chooses which RSU-hours to serve from sleep subject to a single knapsack-like constraint on the traffic they contain, and the CMDP formulation that follows is the sequential, partially informed version of that selection.

\subsection{Complexity Analysis}
\label{sec:complexity}

Problem~\eqref{eq:opt_problem} admits $2^{KT}$ joint schedules, or $2^{KT} = 2^{480} \approx 10^{144}$ RSU-hour assignments for the study corridor ($K = 4$, $T = 120$), and constraint~\eqref{eq:sla_constraint} couples the RSUs, since budget spent by one is unavailable to the others; arrivals are moreover stochastic and only partly predictable at decision time. These features place the problem among constrained multi-agent Markov decision processes, PSPACE-hard in general~\cite{Papadimitriou1987, Zheng2015}; even the single-RSU case requires a linear program over state-action occupancy measures.

A fourth difficulty motivates a step of the solution method rather than the formulation. The standard route to a tractable CMDP solution is Lagrangian relaxation: the constraint is priced by a multiplier and the resulting unconstrained MDP is solved for a deterministic policy, with the multiplier tuned until the constraint is met. Sweeping the multiplier does not trace the achievable frontier continuously. Each value yields a deterministic policy that changes only at discrete breakpoints, so the attainable (energy, violation rate) pairs are the vertices of the convex hull of the frontier and nothing in between, and the gaps between adjacent vertices can be large relative to the budget. In the evaluated corridor the two hull-adjacent policies at I2 achieved violation rates of $1.04\%$ and $0.32\%$, straddling the $\varepsilon = 1\%$ budget with no deterministic policy between them; multiplier tuning alone must reject the first and settle for the second, stranding roughly two thirds of the violation budget and forfeiting the energy it would have bought. Altman~\cite{Altman1999} showed that for a single average-cost constraint the optimal CMDP policy is a randomized mixture of exactly these two hull-adjacent deterministic policies, which recovers the intermediate operating points. A randomized schedule is unattractive for infrastructure operation, so Section~\ref{sec:alg_repair} develops a deterministic construction that plays the same role.

\subsection{Decomposition and Algorithmic Approach}
\label{sec:decomposition}

To make the problem tractable, we adopt a two-level decomposition. At the first level, the corridor constraint~\eqref{eq:sla_constraint} is relaxed by applying it to each RSU separately:
\begin{equation}
\label{eq:per_rsu_constraint}
\frac{\sum_{t=1}^{T} V_k(t)}{\sum_{t=1}^{T} \lambda_k(t)} \leq \varepsilon, \quad \forall\, k.
\end{equation}
The relaxation is conservative, implying~\eqref{eq:sla_constraint} but not implied by it, and reduces the joint problem to $K$ single-RSU subproblems. At the achieved corridor rate of $0.976\%$, $0.024$ percentage points of budget remained unspent, a description of the operating point rather than a bound on the cost of decomposition. The corridor metric remains the primary compliance criterion; per-intersection rates appear in Section~\ref{sec:results_latency}.

At the second level, each single-RSU subproblem is formulated as a constrained MDP. The state of RSU $k$ at time slot $t$ is
\begin{equation}
\label{eq:mdp_state}
\mathbf{x}_k(t) = \bigl(\ell_k(t),\; s_k(t-1),\; h(t),\; d(t)\bigr),
\end{equation}
where $\ell_k(t) \triangleq \mathrm{bin}(\lambda_k(t{-}1)) \in \{1, \ldots, B\}$ is the quantized level of the \emph{previous} hour's arrivals, the most recent observation available when $s_k(t)$ is chosen (the $B$ bins follow the empirical distribution at intersection $k$), $s_k(t-1) \in \mathcal{S}$ is the previous power state, $h(t) \in \{0, \ldots, 23\}$ is the hour of day, and $d(t) \in \{\mathrm{weekday}, \mathrm{weekend}\}$ is the day type. Both $h(t)$ and $d(t)$ are exogenous and known from the calendar. The resulting state space has $|\mathcal{X}_k| = B \times |\mathcal{S}| \times 24 \times 2 = 96B$ states. With $B = 5$ bins (quintile-based discretization), this yields $|\mathcal{X}_k| = 480$ states per RSU, small enough for exact solution via value iteration; value iteration distinguishes 240 of these states, the previous-state axis becoming active only under repair (Section~\ref{sec:alg_mdp}). The day-type covariate is carried through the state and the transition estimates throughout this work; the ablation in Appendix~\ref{app:granularity} shows that it leaves the optimal policy unchanged at this discretization, so it is retained for completeness rather than for the effect it produces.

For the cooperative extension, the state is augmented with the upstream signal:
\begin{equation}
\label{eq:coop_state}
\mathbf{x}_k^{\mathrm{coop}}(t) = \bigl(\ell_k(t),\; s_k(t-1),\; h(t),\; d(t),\; u_{k-1}(t)\bigr),
\end{equation}
where $u_{k-1}(t)$ is the summary received from the upstream neighbor, either the one-bit signal of Eq.~\eqref{eq:upstream_signal} or the bin-level signal of Eq.~\eqref{eq:upstream_signal_bin}. The binary encoding doubles the state space to $|\mathcal{X}_k^{\mathrm{coop}}| = 960$ states and the bin-level encoding multiplies it by $B$, giving $2{,}400$ states, both still tractable for value iteration.

The Lagrangian relaxation of the per-RSU constraint~\eqref{eq:per_rsu_constraint} converts the constrained MDP into an unconstrained MDP with a modified reward function:
\begin{equation}
\label{eq:reward}
R_k(\mathbf{x}, a) = -P(a) \cdot \Delta - \beta_k \cdot V_k(\mathbf{x}, a),
\end{equation}
where the energy term is $\Pawake \Delta$ for the Awake action and the duty-cycle sleep energy $E_k^{\mathrm{sleep}}$ of Eq.~\eqref{eq:sleep_energy} for the Sleep action, $V_k(\mathbf{x}, a)$ is the expected violation count (Eq.~\ref{eq:violations}) given the traffic level encoded in state $\mathbf{x}$, and $\beta_k \geq 0$ is the Lagrange multiplier (violation penalty weight) that trades off energy savings against latency compliance. The target $\beta_k$ is the smallest value for which the resulting unconstrained MDP policy satisfies the $\varepsilon$ violation rate on the training data; it is found via binary search, and the residual gap left by policy quantization is closed by the repair step of Section~\ref{sec:alg_repair}.

%% file: section5_algorithms.tex
This section presents four scheduling algorithms of increasing sophistication for the RSU sleep scheduling problem formulated in Section~\ref{sec:problem_formulation}. The algorithms are organized as a progression: each successive method adds one capability over the previous, enabling a controlled evaluation of the incremental value of adaptivity (Algorithms~2 and~3) and cooperation (Algorithm~4).

\subsection{Algorithm 1: Periodic Fixed-Schedule (PFS)}
\label{sec:alg_periodic}

The periodic fixed-schedule algorithm assigns each RSU a predetermined sleep/wake schedule based on the average hourly traffic profile. The schedule is computed offline, separately for weekday and weekend day types, and does not adapt to real-time traffic conditions.

For each RSU $k$ and day type $d$, the algorithm constructs the schedule as follows. Let $\bar{\lambda}_k^{(d)}(h)$ denote the average arrival rate at RSU $k$ during hour $h$ on day type $d$, computed from the historical traffic data. The 24 hourly slots are sorted in ascending order of $\bar{\lambda}_k^{(d)}(h)$. Starting from the lowest-traffic hour, slots are greedily assigned to Sleep until adding the next slot would cause the expected violation rate to exceed the SLA threshold $\varepsilon$. Formally, the sleep set $\mathcal{H}_k^{(d)} \subseteq \{0, 1, \ldots, 23\}$ is the largest set such that
\begin{equation}
\label{eq:pfs_constraint}
\frac{\sum_{h \in \mathcal{H}_k^{(d)}} V_k(h)}{\sum_{h=0}^{23} \bar{\lambda}_k^{(d)}(h)} \leq \varepsilon,
\end{equation}
where $V_k(h)$ is the expected violation count for hour $h$ computed from Eq.~\eqref{eq:violations} using $\bar{\lambda}_k^{(d)}(h)$. All hours not in $\mathcal{H}_k^{(d)}$ are assigned to the Awake state.

Under the two-state model a single awake state exists, so this algorithm and the three that follow all serve awake hours at $\Pawake$, none handicapped by a costlier awake mode. The schedule is built per RSU and day type, so the SLA binds each RSU against its own arrival mass. A fixed window cannot exploit a quiet hour falling outside it.

\subsection{Algorithm 2: Threshold-Based Reactive (THR)}
\label{sec:alg_threshold}

The threshold-based algorithm makes reactive sleep/wake decisions using the most recent traffic observation. At the beginning of each time slot $t$, RSU $k$ observes the realized arrival count from the previous slot, $\lambda_k(t-1)$, and applies a threshold rule with hysteresis:
\begin{equation}
\label{eq:threshold_rule}
s_k(t) =
\begin{cases}
\mathrm{S}, & \text{if } s_k(t{-}1) = \mathrm{A} \text{ and } \lambda_k(t{-}1) < \lambda_{\mathrm{th}}^{(k)}, \\
\mathrm{A}, & \text{if } s_k(t{-}1) = \mathrm{S} \text{ and } \lambda_k(t{-}1) > 1.5\,\lambda_{\mathrm{th}}^{(k)}, \\
s_k(t{-}1), & \text{otherwise},
\end{cases}
\end{equation}
where $\lambda_{\mathrm{th}}^{(k)}$ is the sleep threshold of RSU $k$ and the factor of 1.5 on the wake-up threshold provides hysteresis to prevent rapid cycling between sleep and awake states during transitional traffic periods.

The threshold is the sole tunable parameter and was calibrated separately for each RSU: for every $k$ a grid search over $[50, 1000]$~veh/hr in steps of 50 selected the value minimizing that RSU's energy subject to its own violation rate satisfying $\varepsilon \leq 0.01$ on the training trace. Per-RSU calibration matters on a heterogeneous corridor: a single corridor-wide threshold is dominated by the aggregate and can leave an individual low-volume intersection above the SLA while the corridor average remains compliant.

The rule adapts to unexpected low-traffic periods that a fixed schedule cannot anticipate, but it conditions on the previous slot's count alone, with no hour-of-day or day-type awareness, and its one-hour observation lag makes it react to traffic changes rather than anticipate them, which costs it during rapid transitions.

\subsection{Algorithm 3: MDP-Based Adaptive (MDP)}
\label{sec:alg_mdp}

The MDP-based algorithm instantiates the per-RSU decomposition of Section~\ref{sec:decomposition}, formulating single-RSU scheduling as a Markov decision process solved by value iteration, yielding the optimal policy for each RSU under the estimated model (Section~\ref{sec:sim_setup} discusses the in-sample nature of this guarantee).

\subsubsection{State, Action, and Transition Model}

The state of RSU $k$ at time slot $t$ is the tuple defined in Eq.~\eqref{eq:mdp_state}, comprising the traffic level $\ell_k(t)$, the previous power state $s_k(t{-}1) \in \{\mathrm{A}, \mathrm{S}\}$, the hour $h(t)$, and the day type $d(t)$. The traffic level $\ell_k(t) \in \{1, \ldots, B\}$ is obtained by discretizing the previous hour's arrival count into $B = 5$ bins, with bin edges set at the quintiles of the empirical arrival rate distribution at intersection $k$. The quintile-based binning ensures that each bin is equally populated in the training data, providing balanced transition probability estimates across all traffic levels.

Under the two-state model the value-iteration optimum is provably independent of the previous-state component: neither the reward nor the transition kernel depends on it, and the unrepaired policies were constant along that axis (the zero-repair policy at I1 has identical slices). The component is retained because the repair edits individual cells and can make the repaired schedule history-dependent, revoking a sleep assignment in one entry state but not the other.

The action set is $\mathcal{A} = \mathcal{S} = \{\mathrm{A}, \mathrm{S}\}$, the action $a_k(t)$ selecting the RSU's power state for the upcoming slot.

The transition probabilities $\Pr[\ell_k(t{+}1) = j \mid \ell_k(t) = i, h(t) = h, d(t) = d]$ capture how the traffic level evolves from one hour to the next. These probabilities are estimated from the 5-day trace by counting, for each $(i, h, d)$ triple, the empirical frequency of transitioning to each bin $j$ at hour $h+1$. Laplace smoothing (adding a count of one to each transition) is applied to avoid zero probabilities, which is necessary given the limited sample size. The sensitivity of the results to the discretization granularity, which implicitly governs the smoothing influence, is evaluated alongside the cooperation results in Section~\ref{sec:results_cooperation}. The hour-of-day component transitions deterministically as $h(t+1) = (h(t) + 1) \bmod 24$, and the day type is fixed within a day and advances at the midnight boundary according to the calendar.

\subsubsection{Reward Function and Penalty Calibration}

The reward function follows the Lagrangian formulation in Eq.~\eqref{eq:reward}:
\begin{equation}
\label{eq:mdp_reward}
R_k(\mathbf{x}, a) = -P(a) \cdot \Delta - \beta_k \cdot V_k(\mathbf{x}, a),
\end{equation}
where the first term is the slot energy of action $a$ and $V_k(\mathbf{x}, a)$ is the expected violation count, both evaluated at the representative arrival rate $\bar{\lambda}(\ell)$ of the traffic bin encoded in $\mathbf{x}$. For $a = \mathrm{A}$ the RSU is awake for the whole slot, so the energy term is $\Pawake \Delta$ and $V_k = 0$. For $a = \mathrm{S}$ the energy term is the duty-cycle sleep energy $E_k^{\mathrm{sleep}}$ of Eq.~\eqref{eq:sleep_energy}, which already includes the wake-up and serving cost of every arrival, and the violation term is $\bar{\lambda}(\ell) \Pviol$ from Eq.~\eqref{eq:violations}. Sleeping is therefore penalized twice over, once through the energy actually spent restoring service and once through the latency exposure it creates.

The penalty weight $\beta_k$ controls the energy-latency tradeoff and is calibrated so that the resulting policy meets the target violation rate $\varepsilon = 0.01$. A higher $\beta_k$ produces a more conservative policy (less sleep, fewer violations); a lower $\beta_k$ produces a more aggressive policy (more sleep, more violations). We determine $\beta_k$ via binary search over the range $[0, 10^8]$, approaching the budget from the feasible side. At each iteration, value iteration is run to convergence, the resulting policy is evaluated on the training trace, and the violation rate is compared to $\varepsilon$. The search isolates the tightest feasible multiplier to a relative tolerance of $10^{-9}$, at which point the achieved rate is limited by the discreteness of the policy space rather than by the search, which is what the repair procedure of Section~\ref{sec:alg_repair} addresses. The invoked structural results are average-cost; the discounted solution ($\gamma = 0.99$, 120 slots) is their computational analogue and exhibits the same hull-vertex behavior (Appendix~\ref{app:stability}).

\subsubsection{Value Iteration}

The policy optimal for the estimated MDP is obtained by solving the Bellman equation
\begin{equation}
\label{eq:bellman}
V^*(\mathbf{x}) = \max_{a \in \mathcal{A}} \left[ R(\mathbf{x}, a) 
+ \gamma \sum_{\mathbf{x}'} \Pr[\mathbf{x}' \mid \mathbf{x}, a] \, 
V^*(\mathbf{x}') \right],
\end{equation}

with discount factor $\gamma = 0.99$. Convergence (defined as $\max_{\mathbf{x}} |V_{n+1}(\mathbf{x}) - V_n(\mathbf{x})| < 10^{-6}$ or policy stability) was reached within 200--400 iterations, well under one second per RSU on commodity hardware. Value iteration is invoked once per step of the $\beta_k$ binary search, so the dominant cost of calibration is the search rather than any single solve.

\subsection{Budget-Repair Portfolio}
\label{sec:alg_repair}

The Lagrangian relaxation yields deterministic policies only at discrete points of the energy-violation frontier. For a single average-cost constraint, the optimal constrained policy is a randomized mixture of the two deterministic policies adjacent to the budget~\cite{Altman1999}; a randomized schedule, however, is undesirable for infrastructure operation. We therefore construct a deterministic analogue of the mixture by repair: starting from the frontier policy that overspends the budget, individual sleep decisions are revoked one at a time until the budget is met; a pairwise local search then exchanges sleep assignments to recover energy while maintaining feasibility. Which decision is revoked next is set by a greedy selection rule, and the outcome is sensitive to that choice, so the procedure is run under a portfolio of five deterministic selection rules and the best feasible result is kept. The repaired policy is deterministic and was feasible in every evaluated instance, with the portfolio selection restricted to rules that reached feasibility, and it attained the budget within $0.036$ percentage points at every RSU.

Algorithm~\ref{alg:repair} states the procedure. The two inputs $\pi_{\mathrm{high}}$ and $\pi_{\mathrm{low}}$ are the hull-adjacent policies produced by the $\beta_k$ binary search, the last infeasible policy and the first feasible one. Repair is skipped when $\pi_{\mathrm{low}}$ already sits within a small slack of the budget, since there is then nothing to recover. Every candidate edit is scored by replaying the policy on the trace rather than by a surrogate, so the cascade effects that a single changed decision induces through the previous-state component of the state are captured exactly. The candidate set for the local search is restricted to policy cells that the replay actually reaches, which prunes only edits that provably change nothing.

\begin{algorithm}[tb]
{\footnotesize\caption{Budget-Repair Portfolio}\label{alg:repair}}
\begin{algorithmic}[1]
\renewcommand{\algorithmicrequire}{\textbf{Input:}}
\renewcommand{\algorithmicensure}{\textbf{Output:}}
\REQUIRE Hull-adjacent policies $\pi_{\mathrm{high}}$ (rate $> \varepsilon$) and $\pi_{\mathrm{low}}$ (rate $\leq \varepsilon$), trace, budget $\varepsilon$, slack $\delta = 0.02$ percentage points, selection rules $\mathcal{O}$
\ENSURE Deterministic feasible policy $\pi^\star$
\medskip
\STATE \textbf{if} $\varepsilon - \mathrm{rate}(\pi_{\mathrm{low}}) \leq \delta$ \textbf{then return} $\pi_{\mathrm{low}}$ \quad // budget already tight
\STATE $\mathcal{P} \gets \emptyset$
\FOR{each greedy selection rule $o \in \mathcal{O}$}
    \STATE $\pi \gets \pi_{\mathrm{high}}$ \quad // greedy demotion
    \WHILE{$\mathrm{rate}(\pi) > \varepsilon$}
        \STATE $c^\star \gets \arg\min_{c} \bigl\{ \kappa_o(c) : \pi(c) = \mathrm{S},\; \Delta V(c) > 0 \bigr\}$, ties broken by cell index
        \STATE $\pi(c^\star) \gets \mathrm{A}$; re-evaluate $\pi$ on the trace
    \ENDWHILE
    \STATE apply pairwise local search to $\pi$
    \STATE $\mathcal{P} \gets \mathcal{P} \cup \{(\pi, E(\pi), o)\}$
\ENDFOR
\STATE \textbf{return} $\pi^\star$, the least-energy feasible element of $\mathcal{P}$, ties broken by rule name
\end{algorithmic}
\end{algorithm}

The five selection rules are not tie-breakings of one criterion but five distinct greedy objectives, each a different $\kappa_o$ over the candidate cells: $\kappa_{\mathrm{ratio}}(c) = -\Delta V(c)/\Delta E(c)$, $\kappa_{\mathrm{energy}}(c) = \Delta E(c)$, $\kappa_{\mathrm{violations}}(c) = -\Delta V(c)$, and $\kappa_{\mathrm{index\text{-}asc}}$ and $\kappa_{\mathrm{index\text{-}desc}}$, which order the cells lexicographically by index in ascending and descending order and ignore both $\Delta V$ and $\Delta E$. Here $\Delta V(c)$ is the violation count removed and $\Delta E(c)$ the energy saving forgone by revoking the sleep assignment at cell $c$, both measured by exact replay. Only candidates with $\Delta V(c) > 0$ are admitted, so every accepted flip strictly reduces violations and the loop terminates.

The procedure is deterministic end to end: the greedy selection rules, the local search, and the final portfolio selection all resolve ties by fixed rules. It applies unchanged to the independent MDP of Section~\ref{sec:alg_mdp} and to the cooperative variants of Section~\ref{sec:alg_cooperative}, keeping the comparison free of any difference in solver strength.

\subsection{Algorithm 4: Corridor-Aware Cooperative (COOP)}
\label{sec:alg_cooperative}

The cooperative algorithm extends the single-RSU MDP with the upstream traffic-state summary defined in Section~\ref{sec:cooperative_model}. Its claim is narrow and stated here for precision: the summary is an additional observation that may sharpen the downstream RSU's forecast of next-hour traffic, and nothing in the mechanism provides advance warning of an individual vehicle or enables a wake-up earlier than the arrival that triggers it. Whether the sharper forecast translates into measurable energy savings is settled empirically in Section~\ref{sec:results_cooperation}.

\subsubsection{State Augmentation}

For downstream RSUs ($k \geq 2$), the state is augmented with the upstream signal as defined in Eq.~\eqref{eq:coop_state}:
\[
\mathbf{x}_k^{\mathrm{coop}}(t) = \bigl(\ell_k(t),\; s_k(t{-}1),\; h(t),\; d(t),\; u_{k-1}(t)\bigr).
\]
Two encodings of $u_{k-1}(t)$ are evaluated, both defined in Section~\ref{sec:cooperative_model}. The default is the one-bit signal of Eq.~\eqref{eq:upstream_signal}, which doubles the state space to $|\mathcal{X}_k^{\mathrm{coop}}| = 5 \times 2 \times 24 \times 2 \times 2 = 960$ states. The pre-specified variant COOP-BIN uses the bin-level signal of Eq.~\eqref{eq:upstream_signal_bin}, giving $5 \times 2 \times 24 \times 2 \times 5 = 2{,}400$ states. The head-of-corridor RSU ($k = 1$) has no upstream neighbor and uses the 480-state MDP of Algorithm~3 in both variants.

\subsubsection{Modified Transition and Reward Model}

The upstream signal modifies the MDP in two ways. First, the traffic-level transition probabilities are reweighted and renormalized,
\begin{equation}
\label{eq:coop_transition}
\Pr\nolimits^{\mathrm{coop}}[\ell' \!=\! j \mid \ell, h, d, u]
= \frac{w_j(u)\,\Pr[\ell' \!=\! j \mid \ell, h, d]}
       {\sum_{i=1}^{B} w_i(u)\,\Pr[\ell' \!=\! i \mid \ell, h, d]},
\quad
w_j(u) = \begin{cases} 1 + \rho_{k-1,k}\, s_j, & u = \mathrm{High},\\[2pt] 1, & u = \mathrm{Low},\end{cases}
\end{equation}
where $s_j = -1$, $0$, $+1$ for bins below, at, and above the middle bin $\lfloor B/2 \rfloor$, so a High signal moves mass from the low bins to the high ones while a Low signal leaves the estimated transitions unchanged; the weighted products are floored at a small positive constant before renormalization to keep the chain irreducible. Second, the Sleep reward is evaluated at the effective arrival rate of Eq.~\eqref{eq:lambda_eff}, which scales $\lambda$ by $1 \pm \rho_{k-1,k}\alpha$ with $\alpha = 0.30$. Both encodings defined in Section~\ref{sec:cooperative_model} enter this construction unchanged, the bin-level variant replacing the binary offset with a centered deviation across the $B$ upstream bins and drawing its transition model directly from the trace conditioned on that bin. Neither encoding provides advance warning of an individual vehicle, and no preemptive wake-up is claimed; the only mechanism is a possibly sharper next-hour forecast.

\subsubsection{Solution and Complexity}

The cooperative MDP is solved by the same value iteration procedure as Algorithm~3 (Eq.~\ref{eq:bellman}), with the state space expanded as above, and is then passed through the identical budget-repair portfolio of Section~\ref{sec:alg_repair}. The penalty weight $\beta_k$ is re-calibrated independently for each downstream RSU, since the upstream information changes the energy-violation tradeoff. In the value iteration, the next slot's upstream signal is marginalized under its empirical hourly frequency, treated as conditionally independent of the traffic bin given the hour; the signal therefore informs the current decision but carries no persistence into the value function. Using the same calibration and repair pipeline for Algorithms~3 and~4 is deliberate: any difference in reported performance is then attributable to the information available to the policy, not to the solver.

Measured end to end, the offline pipeline took at most $17.3$~s per RSU and $54.4$~s for the four-RSU corridor, dominated by the $\beta_k$ search and the five-rule repair portfolio. At runtime each RSU performs an $O(1)$ table lookup and exchanges one message per slot, so corridor overhead scales linearly with $K$.

Offline, and once per RSU, the bin edges and transition probabilities are estimated, the value iteration of Eq.~\eqref{eq:bellman} is solved, $\beta_k$ is calibrated by binary search, and the resulting hull-adjacent pair is passed to Algorithm~\ref{alg:repair}. Online, each RSU observes its traffic bin, previous state, hour, and day type, receives the upstream summary if it has a neighbor, looks up its action, and forwards its own summary downstream.

%% file: section6_results.tex
This section evaluates the four scheduling algorithms on the 5-day traffic trace from the four-intersection corridor described in Section~\ref{sec:corridor_model}.

\subsection{Simulation Setup}
\label{sec:sim_setup}

The simulation replayed the 5-day hourly traffic trace (762,050 total vehicles across 4 intersections and 120 time slots) through each scheduling algorithm. The RSU power parameters follow Table~\ref{tab:power_params}, with the latency constraint $\Lmax = 100$~ms and violation threshold $\varepsilon = 0.01$. The always-on (AO) baseline, all RSUs Awake continuously, is the reference for energy savings, consuming 3,628,800~J (1,008~Wh) per RSU over the 5-day period.

Every algorithm served its awake hours in the single Awake state (Section~\ref{sec:power_model}). All four were calibrated per RSU against that RSU's own violation rate, and the two MDP-based algorithms passed through the identical budget-repair portfolio of Section~\ref{sec:alg_repair}, so any difference between them reflected the information available to the policy rather than the strength of the solver.

For the MDP and cooperative algorithms, the traffic arrival rates were discretized into $B = 5$ bins per RSU (quintile-based), yielding 480-state and 960-state MDPs respectively. Transition probabilities were estimated from the same 5-day trace used for evaluation. This overlap between training and evaluation data means the reported MDP and cooperative savings represent upper bounds on deployable performance; the magnitude of this optimism cannot be estimated from a five-day trace, but the stability of retrained policies across a twelvefold traffic range (Appendix~\ref{app:sweeps}) indicates the policies are not brittle to the specific trace. Calibration targets the empirical trace; under the estimated kernel the repaired policies' stationary rates can exceed $\varepsilon$ (Appendix~\ref{app:stability} reports 1.08\% at I3, 1.70\% at I4, and the LP comparison at matched stationary compliance). A statistically meaningful train-test split is not available from five days, since holding out one weekend day leaves a single weekend training day; the traffic scaling analysis in Appendix~\ref{app:sweeps} provides indirect evidence of generalizability. All results are deterministic and fully reproducible: a fixed trace with no stochastic sampling.

\subsection{Energy Savings Comparison}
\label{sec:results_energy}

The left panel of Table~\ref{tab:energy_results} presents the 5-day percentage savings for each algorithm. The cooperative algorithm reached 39.67\% corridor-level energy savings over the always-on baseline, cutting total corridor energy from 14,515,200~J to 8,757,605~J. The independent MDP reached 39.46\% and the best simple baseline (Periodic) 36.26\%, so the adaptive methods improved on the simple ones by 3.20 percentage points and differed from one another by 0.21.

\begin{table}[tb]
\centering
\footnotesize
\caption{Corridor and per-intersection outcomes over the 5-day trace: energy savings against the always-on baseline, and latency violation rates, all within $\varepsilon = 1\%$}
\label{tab:energy_results}
\footnotesize
\setlength{\tabcolsep}{4pt}
\begin{tabular}{@{}lrrrrr@{\hspace{7pt}}|@{\hspace{7pt}}rrrrr@{}}
\toprule
& \multicolumn{5}{c}{\textit{Energy savings} (\%)} & \multicolumn{5}{c}{\textit{Latency violation rate} (\%)} \\
\cmidrule(lr){2-6} \cmidrule(lr){7-11}
Algorithm & I1 & I2 & I3 & I4 & Corridor & I1 & I2 & I3 & I4 & Corridor \\
\midrule
Periodic & 36.30 & 36.18 & 33.79 & 38.75 & 36.26 & 0.834 & 0.835 & 0.880 & 0.921 & 0.865 \\
Threshold & 37.06 & 37.70 & 31.31 & 38.73 & 36.20 & 0.940 & 0.991 & 0.926 & 0.963 & 0.957 \\
MDP & 39.48 & 39.35 & 36.19 & 42.81 & 39.46 & 0.984 & 0.973 & 0.975 & 0.973 & 0.976 \\
Cooperative & 39.48 & 39.35 & 36.19 & 43.63 & 39.67 & 0.984 & 0.968 & 0.964 & 0.969 & 0.971 \\
\bottomrule
\end{tabular}
\end{table}

The four algorithms differed only in \emph{which} hours they placed in Sleep and therefore in how much traffic they exposed to restoration delay. Remark~\ref{lem:budget} makes that comparison precise: the SLA is a budget permitting at most $\varepsilon / \Pviol = 14.29\%$ of arrivals to fall in slept hours, and an algorithm's energy saving is governed by how completely it spends that budget on the cheapest available hours. Table~\ref{tab:slept_share} in Appendix~\ref{app:granularity} reports the share each algorithm actually consumed.

The two simple baselines left budget unused. The periodic schedule spent only $12.35\%$ of the corridor budget, since it commits to a fixed block of hours per day type and cannot exploit a quiet hour outside it; the threshold rule reached $13.67\%$ by reacting to observed dips, but its one-hour lag and hysteresis left it short. The two MDP-based algorithms rode the budget almost exactly, at $13.94\%$ and $13.87\%$, every RSU under both within $0.51$ percentage points of the $14.29\%$ ceiling. That is the whole of their advantage: the $3.20$ percentage point improvement over the periodic baseline was bought by converting unused violation budget into slept hours, not by a cheaper awake state. Near-identical savings do not imply near-identical schedules: the MDP reached $39.48\%$ at I1 and $39.35\%$ at I2 from the same 49 sleep hours but different placements (Fig.~\ref{fig:gantt_weekday}), I1 requiring no repair at all and I2 one demotion, while I4 required 28.

Table~\ref{tab:state_hours} in Appendix~\ref{app:granularity} gives the hour counts: the MDP-based algorithms slept 45 to 54 hours per RSU against 39 to 48 for the baselines, the extra hours being low-traffic ones that fit inside the remaining budget. Cooperation added one slept hour at I4 (54 against 53) and matched the MDP elsewhere, for a corridor total of 197 hours against 196.

The weekday-weekend split was pronounced. Lower weekend volumes (116,528 against 176,332 daily corridor vehicles) place more hours within reach of the budget, so the cooperative algorithm saved 51.76\% on weekends against 31.26\% on weekdays and the MDP 50.76\% against 31.93\%, while the periodic schedule stayed nearly flat (36.37\% and 36.18\%), its fixed block being sized for the average day. The realized schedules differed in structure as well as extent (Appendix~\ref{app:granularity}, Fig.~\ref{fig:gantt_weekday}).

\subsection{Latency Compliance Analysis}
\label{sec:results_latency}

The right panel of Table~\ref{tab:energy_results} reports the per-intersection and corridor-level latency violation rates. All four algorithms satisfied the corridor-level SLA constraint ($\varepsilon = 0.01$), with corridor rates from 0.8646\% (Periodic) to 0.9760\% (MDP), and every algorithm was also compliant at every individual intersection, the threshold rule at I4 by the narrowest margin (0.963\%).

A single corridor-aggregate threshold, by contrast, breaches $\varepsilon$ at I4 (1.30\%) while remaining corridor-compliant (0.95\%), which is why $\lambda_{\mathrm{th}}$ is calibrated per RSU; the per-RSU choice costs the sleep hours the aggregate threshold had been buying at I4's expense.

The two MDP-based algorithms sat closest to the constraint, at 0.976\% and 0.971\% corridor-wide, the intended effect of the repair: the repaired policies exhaust the budget rather than strand it. Their per-RSU rates spanned 0.964\% to 0.984\%, all within $0.036$ percentage points of $\varepsilon$, and the two differed by at most $0.011$ percentage points at every downstream RSU. That places them at matched operating points, which is what makes the comparison in Section~\ref{sec:results_cooperation} meaningful.

\subsection{Benefit of Corridor Cooperation}
\label{sec:results_cooperation}

Corridor cooperation delivered its benefit in inverse proportion to the coupling its design relies on. At matched compliance, the cooperative policy matched the independent MDP to within $0.007$ percentage points at the two most strongly correlated pairs (I2, $\rho = 0.977$; I3, $\rho = 0.947$), and only the weakest-coupled pair (I4, $\rho = 0.848$) showed a gain, of $0.820$ percentage points, diluting to $0.208$ percentage points at corridor level ($39.67\%$ against $39.46\%$). This ordering is the inverse of the design premise that stronger correlation makes the upstream signal more valuable, and it is the ordering that redundancy predicts: a strongly correlated neighbor is a near-copy of what the downstream RSU already observes; that ordering is an observation at the full five-day sample, and under three- and four-day training prefixes the corridor gain remains within 0.6 percentage points of zero but the per-RSU pattern varies (Appendix~\ref{app:stability}). The I4 gain was consistent in sign across both signal encodings ($+0.820$ binary, $+0.826$ bin-level) and all five greedy selection rules, but it is modest, sits at the lower edge of its solver-sensitivity band (Appendix~\ref{app:stability}), and is confounded with I4's distinctive traffic profile as the corridor's only three-way intersection, so it cannot be attributed to the signal mechanism alone. With a single pair in the weak-coupling regime, a margin admitted by weak coupling cannot be separated from one admitted by I4's three-way geometry; the ordering claim therefore rests on the two strongly coupled pairs, where the absence of any gain is unconfounded and consistent across encodings, selection rules, and modulation strengths, with the I4 sliver read as consistent with the ordering rather than as evidence for it. It also flowed entirely through the reward-modulation channel: it vanished at $\alpha = 0$ and required no measurable predictive information in the signal, consistent with the modulation acting as a risk adjustment at marginal sleep decisions rather than as prediction. Across the sweep, the corridor gain never exceeded $0.21$ percentage points at any $\alpha \in [0, 0.5]$ and turned negative at $\alpha \geq 0.4$, with $\alpha = 0.30$ the pre-registered default (Appendix~\ref{app:alpha}). Table~\ref{tab:cooperation_gain} in Appendix~\ref{app:coopbin} reports the per-RSU and corridor gains for both encodings, measured against the independent MDP at the matched operating points established in Section~\ref{sec:results_latency}. The upstream signal enters the value function only through the current slot (Section~\ref{sec:alg_cooperative}), so the mechanism cannot accumulate value across slots.

The mechanism of both the zeros and the sliver at I4 is identifiable in the data: only where the correlation weakens does any margin appear. First, the binary upstream signal's marginal frequency was exactly $0.40$ at every hour of the day (the median is taken within day type, so one of three weekdays and one of two weekend days exceed it), so the augmentation added no hour-specific structure, only a within-hour day ranking. Second, the richer five-level encoding, although less destructive, still degraded out-of-sample prediction of the downstream traffic bin at all three downstream RSUs (leave-one-out log-loss increases of $0.03$--$0.09$), because 119 one-step training transitions per RSU spread over the enlarged conditioning space leave roughly three quarters of the predictive mass to the smoothing prior. Third, and most fundamentally, the information the signal could carry is largely redundant: conditional on the downstream RSU's own traffic bin and the hour of day, the upstream signal added no statistically detectable information about the next-hour downstream bin (permutation $p = 0.66$--$0.98$). With 119 transitions distributed over 50 to 56 conditioning strata, this test has limited statistical power, so the result establishes the absence of a detectable effect rather than the absence of any effect; the out-of-sample degradation, however, is a positive finding that conditioning on the signal harms prediction. The spatial correlation that motivates cooperation is precisely what makes the upstream observation a near-copy of information the downstream RSU already possesses.

The premise of corridor cooperation is not thereby refuted. The continuous upstream rate retained a positive partial correlation with the next-hour downstream rate after controlling for local rate and hour ($+0.570$, $+0.314$, and $+0.211$ at I2--I4, $p \leq 0.037$), so predictive information exists in the corridor; it is the hourly time scale and the quantized encoding that delivered no more than a marginal fraction of it to the scheduler. Cooperation should therefore be re-examined at scheduling intervals commensurate with the inter-RSU travel time of 1--2 minutes, where upstream observations lead downstream arrivals rather than merely summarizing a shared diurnal pattern. At the hourly granularity of practical schedule computation with the present data, the added signaling and state complexity of cooperation was not justified by a corridor gain of at most $0.21$ percentage points.

Policy granularity mattered far less once the budget was repaired: savings varied by only $1.64$, $0.83$, and $0.83$ percentage points across $B \in \{3, 4, 5, 7, 10, 15, 20\}$ at I2, I3, and I4, with the variation at I2 confined almost entirely to $B = 3$. The same pipeline without the repair step swung by roughly $14$ percentage points over the same range and showed a cooperation gain that appeared to peak at $B = 5$; both were quantization artifacts and neither survives repair (Appendix~\ref{app:granularity}).

\subsection{Sensitivity Analysis}
\label{sec:results_sensitivity}

Four parameters were swept, each with the policy recalibrated at every point: the latency SLA, the hardware wake-up delay, the traffic volume, and the RSU power class.

\subsubsection{Latency SLA}

Corridor energy savings depended sharply on the latency budget $\Lmax$ (Appendix~\ref{app:sweeps}, Fig.~\ref{fig:sensitivity_sla}). Savings rose from $11.38\%$ at $\Lmax = 50$~ms through $18.48\%$ at 75~ms, $39.67\%$ at the 3GPP operating point of 100~ms, and $67.15\%$ at 105~ms, then jumped to a plateau of $93.77\%$ at 107~ms and stayed there for every larger budget. The knee sits exactly at $\Lmax = \twake + \Ltx + T_{\mathrm{WSA}} = 107$~ms, which by Eq.~\eqref{eq:Pviol} is the point where $\Pviol$ reaches zero: beyond it no arrival at a sleeping RSU can exceed the budget, the violation constraint ceases to bind, and every hour becomes eligible for sleep. Below the knee the constraint is active and the scheduler must choose which hours to spend it on.

The plateau is $93.77\%$ rather than $100\%$ because sleeping is not free even when unconstrained. Under the duty-cycle model a slept hour still wakes the RSU on each arrival and serves it at $\Pawake$ for $\Tup = 107$~ms per cycle (Eq.~\eqref{eq:sleep_energy}), holding it near $6\%$ of an awake hour (Appendix~\ref{app:sweeps}) and separating this figure from the $\sim$98\% a model omitting it would report. The optimization is non-trivial only when $\Lmax < 107$~ms; the TS~22.185 budget of 100~ms falls just inside that boundary, and the tighter TS~22.186 advanced-driving requirements would bind harder still.

The remaining three sweeps, over wake-up delay, traffic volume, and RSU power class, are detailed in Appendix~\ref{app:sweeps}: savings degraded smoothly with wake-up delay and the portfolio solver returned a feasible schedule at every tested value, with no collapse; retrained policies held near-constant savings across a twelvefold volume range, the burden of adaptation falling on the discretization rather than the policy structure; and percentage savings were nearly invariant to power class.

\subsection{Carbon Footprint and Deployment Cost Reduction}
\label{sec:results_carbon}

The energy savings translate into annual carbon dioxide reduction at a factor of 0.55~kg~CO$_2$/kWh, a representative recent value for Kuwait's grid (0.49--0.56~kg~CO$_2$/kWh)~\cite{Ember2025}. The cooperative algorithm saved an average of 16.05~kg~CO$_2$ per RSU per year, 64.2~kg for the 4-RSU study corridor, against 15.97~kg and 63.9~kg for the independent MDP. Extrapolated to a hypothetical 200-RSU deployment across Kuwait City, a moderate scale for a smart-city V2I network, the adaptive algorithms reduce emissions by 3.19 to 3.21~tonnes~CO$_2$ per year against 2.93~tonnes for the periodic baseline. Table~\ref{tab:carbon} in Appendix~\ref{app:sweeps} gives the per-algorithm breakdown in both energy and carbon units.

These figures are modest in absolute terms, are a lower bound, and are illustrative: the 200-RSU scale is an assumption and the factor an average rather than marginal one. The 8.4~W MK5-class RSU sits at the low end of the commercial power range, and dual-mode RSUs (15--25~W) yield proportionally greater absolute savings at a similar percentage, as the power-class sweep showed.

%% file: conclusion.tex
This paper studied sleep scheduling for roadside units under a latency service constraint, on five days of measured traffic from four intersections in Kuwait City. A Sleep-scheduled hour was modeled as a duty cycle charging the transition and serving energy of every restoration it induces. Since awake hours are violation-free and every arrival at a sleeping unit carries the same violation probability, the SLA is a budget permitting at most 14.3\% of arrivals in slept hours. Lagrangian calibration reaches only discrete points of the energy-violation frontier, so a deterministic budget-repair portfolio was introduced to drive policies to that boundary. The repaired policies saved 39.5\% to 39.7\% of corridor energy at 99\% latency compliance, against 36.3\% for a periodic schedule under the same constraint; because policies are trained and evaluated on the same trace, these are upper bounds on deployable performance (Section~\ref{sec:sim_setup}).

Corridor cooperation delivered at most a marginal improvement over independent per-RSU scheduling: zero to within 0.007 percentage points at the two most strongly correlated pairs, 0.8 percentage points at the weakest, and 0.2 at corridor level, an ordering, at the full five-day sample, inverse to the design premise and consistent with the redundancy of a strongly correlated neighbor's signal. The boundary is one of timescale rather than of principle, and its direct test is to collect sub-hourly counts and repeat the comparison at that granularity. The design-space analysis located a knee at $\Lmax = 107$~ms, where the restoration interval fits inside the latency budget, and annualizes the reduction at 3.2~tonnes CO$_2$ per year for an illustrative 200-RSU deployment.
\label{pg:viiend}

%% file: appendix.tex
\section{Empirical Basis of the Upstream Signal}
\label{app:signal}

The corridor exhibits strong contemporaneous coupling. Computed on the 119-hour lag-aligned subset used throughout this appendix (the final hour of the trace has no successor), the Pearson correlations between adjacent hourly volume series are $0.977$, $0.950$, and $0.852$ for the pairs I1$\to$I2, I2$\to$I3, and I3$\to$I4; the third-decimal difference from Eq.~\eqref{eq:correlations}, which uses all 120 hours, reflects that single dropped observation. The quantity relevant to a slot-$t+1$ decision is the lagged coupling, $\mathrm{corr}(\lambda_k(t), \lambda_{k+1}(t+1))$, which is $0.954$, $0.893$, and $0.736$ for the same pairs. Because both series share a pronounced diurnal cycle, part of this lagged association is explained by the hour of day and by the downstream RSU's own current traffic. These are the measurements on which the cooperative signal design in Section~\ref{sec:cooperative_model} rests, and against which the cooperation result of Section~\ref{sec:results_cooperation} should be read.

\section{Budget-Repair Solver: Optimality and Stability}
\label{app:stability}

The state space is small enough that the exact constrained solution is available: the linear program over state-action occupancy measures cited in Section~\ref{sec:rw_cmdp} has 960 variables at $|\mathcal{X}_k| = 480$ and two actions. Solving it per RSU bounds what the deterministic repair gives up. Because the $\beta$ search calibrates against the trace while the program constrains the stationary rate of the estimated kernel, both objects are compared at matched stationary compliance: the program is solved at each repaired policy's own stationary violation rate, so the energy difference is an optimality gap rather than a difference of operating points. The trace is treated as ground truth and the kernel as a planning device: its predictive mass is majority prior at this sample size (Appendix~\ref{app:coopbin}), and the stationary rates of the repaired policies reach 1.08\% at I3 and 1.70\% at I4 even where the trace rates sit below $\varepsilon$. Table~\ref{tab:lpgap} reports the comparison.

The optimal occupancy measure mixes actions in exactly one state of 480 at every RSU, confirming empirically the single-constraint structure of~\cite{beutler1985, Altman1999}: the randomized optimum is one state away from deterministic, which is precisely the situation a deterministic repair step is designed for. The repaired policies attain the LP optimum exactly at I1 and fall within 1.9 percentage points of it elsewhere (mean 1.04).

\begin{table}[t]
\centering
\footnotesize
\caption{Exact constrained optimum against the repaired deterministic policy, both evaluated under the stationary distribution of the estimated kernel at matched compliance}
\label{tab:lpgap}
\small
\setlength{\tabcolsep}{4pt}
\begin{tabular}{@{}lrrr@{}}
\toprule
RSU & LP optimum (\%) & Repaired (\%) & Gap (pp) \\
\midrule
I1 & 38.30 & 38.30 & 0.00 \\
I2 & 38.38 & 37.84 & 0.54 \\
I3 & 37.93 & 36.20 & 1.73 \\
I4 & 53.92 & 52.01 & 1.91 \\
\bottomrule
\end{tabular}
\end{table}

A single ratio-greedy rule with local search matches the portfolio to within 0.01 percentage points on every algorithm and RSU (Table~\ref{tab:ratioport}); the portfolio is retained as insurance against pathological instances rather than as a source of performance. The wide selection-rule bands of Table~\ref{tab:stability} concern the cooperation gain, a difference of two independently repaired policies, not either policy's savings.

\begin{table}[t]
\centering
\footnotesize
\caption{Single ratio-greedy rule with local search against the five-rule portfolio, corridor-component savings}
\label{tab:ratioport}
\small
\setlength{\tabcolsep}{4pt}
\begin{tabular}{@{}llrrr@{}}
\toprule
Algorithm & RSU & Ratio only (\%) & Portfolio (\%) & Delta (pp) \\
\midrule
MDP & I1 & 39.485 & 39.485 & 0.000 \\
MDP & I2 & 39.343 & 39.349 & 0.006 \\
MDP & I3 & 36.178 & 36.186 & 0.008 \\
MDP & I4 & 42.805 & 42.814 & 0.010 \\
\midrule
COOP & I2 & 39.351 & 39.353 & 0.001 \\
COOP & I3 & 36.191 & 36.192 & 0.001 \\
COOP & I4 & 43.634 & 43.634 & 0.000 \\
\bottomrule
\end{tabular}
\end{table}

The budget-repair procedure of Section~\ref{sec:alg_repair} chooses the next demotion by a greedy selection rule, and the locked solver evaluates all five rules and keeps the best feasible result. Running each rule in isolation quantifies how much that arbitrary choice would otherwise move the outcome. Table~\ref{tab:stability} reports the cooperation gain obtained under each rule, with the pairwise local search enabled throughout. The reported per-RSU gains of Section~\ref{sec:results_cooperation} lie inside these bands, and at I4 the band spans an order of magnitude. Because the portfolio selects the minimum-energy feasible policy for each algorithm independently, the reported gains at I2 and I4 sit at the lower edge of their selection-rule bands and the I3 gain mid-band; the band's far end reflects weaker solves, so the reported figures are the conservative estimates.

\begin{table}[t]
\centering
\footnotesize
\caption{Cooperation gain (percentage points vs.\ the independent MDP) under each greedy selection rule, local search enabled}
\label{tab:stability}
\small
\setlength{\tabcolsep}{4pt}
\begin{tabular}{@{}lrrr@{}}
\toprule
RSU & Minimum & Maximum & Range \\
\midrule
I2 & $+0.004$ & $+0.009$ & 0.005 \\
I3 & $+0.002$ & $+0.019$ & 0.017 \\
I4 & $+0.820$ & $+9.018$ & 8.198 \\
\bottomrule
\end{tabular}
\end{table}

Refitting on calendar prefixes of three and four days and evaluating on the full trace, the corridor cooperation gain remains within 0.6 percentage points of zero at every training size (Table~\ref{tab:nullstab}). The per-RSU pattern is not stable: the I3 gain ranges from $-3.2$ to $+0.007$ percentage points across training sizes, so the ordering of per-RSU gains reported at five days should be read as an observation at this sample, not a stable property. The prefixes are a stability curve rather than a validation split, since the evaluation trace contains the fitting days.

\begin{table}[t]
\centering
\footnotesize
\caption{Cooperation gain (percentage points) with the model fitted on calendar prefixes and evaluated on the full five-day trace}
\label{tab:nullstab}
\small
\setlength{\tabcolsep}{4pt}
\begin{tabular}{@{}lrrrr@{}}
\toprule
Training days & I2 & I3 & I4 & Corridor \\
\midrule
3 & $-0.002$ & $-3.207$ & $+0.825$ & $-0.596$ \\
4 & $+0.000$ & $-2.505$ & $+1.583$ & $-0.231$ \\
5 & $+0.004$ & $+0.007$ & $+0.820$ & $+0.208$ \\
\bottomrule
\end{tabular}
\end{table}

\section{Modulation-Strength Sweep}
\label{app:alpha}

The cooperative modulation parameter $\alpha$ of Eq.~\eqref{eq:lambda_eff} scales how far an upstream summary shifts the downstream traffic forecast. It is a design parameter rather than a measured quantity, so its influence was assessed by sweeping it with the policy fully recalibrated at each value. Within the update, the measured correlation scales the modulation, so at $\rho_{1,2} = 0.977$ (I1$\to$I2) the upstream summary moves the downstream forecast further than at $\rho_{3,4} = 0.848$ (I3$\to$I4). Table~\ref{tab:alpha} reports corridor savings and violation rates. Savings vary by 0.82 percentage points across the range and compliance is held throughout. The corridor gain never exceeded $+0.21$ percentage points anywhere in the sweep, itself at the lower edge of the I4 selection-rule band (Appendix~\ref{app:stability}), and turned negative for $\alpha \geq 0.4$; $\alpha = 0.30$ is the pre-registered default.

\begin{table}[t]
\centering
\footnotesize
\caption{Corridor energy savings and violation rate against the upstream modulation strength $\alpha$, with full recalibration at each value}
\label{tab:alpha}
\small
\setlength{\tabcolsep}{4pt}
\begin{tabular}{@{}lrrr@{}}
\toprule
$\alpha$ & Savings (\%) & Violation rate (\%) & Gain vs.\ MDP (pp) \\
\midrule
0.0 & 39.459 & 0.976 & $+0.000$ \\
0.1 & 39.666 & 0.972 & $+0.208$ \\
0.2 & 39.666 & 0.972 & $+0.208$ \\
0.3 & 39.666 & 0.971 & $+0.208$ \\
0.4 & 39.049 & 0.977 & $-0.409$ \\
0.5 & 38.846 & 0.974 & $-0.612$ \\
\bottomrule
\end{tabular}
\end{table}

\section{Bin-Level Signal Variant}
\label{app:coopbin}

The default cooperative algorithm transmits the one-bit upstream signal of Eq.~\eqref{eq:upstream_signal}. A pre-specified variant, COOP-BIN, transmits instead the upstream RSU's five-level quantile bin, Eq.~\eqref{eq:upstream_signal_bin}, at a cost of three bits per slot and a fivefold larger conditioning space for the transition estimates. Table~\ref{tab:cooperation_gain} compares both encodings against the independent MDP at matched compliance. Because the enlarged space is estimated from the same trace, the Laplace prior supplies 72.2\% to 73.5\% of the predictive mass at the cells the replay visits. The bin-level encoding's loss at I3 ($-0.83$ percentage points, against $+0.007$ for the binary encoding at the same RSU) is read as a conditioning effect rather than an information effect: the enlarged state space spreads the same 119 transitions over five times as many conditioning cells, leaving roughly three quarters of the predictive mass to the smoothing prior and degrading both the transition estimates and the repair search. The base transition estimates are comparably prior-dominated (70\% of predictive mass at visited cells, against 72.2\% to 73.5\% for the bin-level variant), so the smoothing burden is a property of the sample size, not of the state augmentation; the binary cooperative variant reweights the base estimates analytically and inherits the same share.

\begin{table}[t]
\centering
\footnotesize
\caption{Cooperation gain at matched compliance (percentage points vs.\ the independent MDP) for both signal encodings, with the independent MDP's absolute savings for reference}
\label{tab:cooperation_gain}
\small
\setlength{\tabcolsep}{3pt}
\begin{tabular}{@{}lcrrr@{}}
\toprule
RSU & Upstream ($\rho$) & MDP & Binary $u$ & Bin-level $u$ \\
\midrule
I1 & --- (head) & 39.48\% & 0.000 & 0.000 \\
I2 & I1 ($\rho = 0.977$) & 39.35\% & $+0.004$ & $+0.004$ \\
I3 & I2 ($\rho = 0.947$) & 36.19\% & $+0.007$ & $-0.829$ \\
I4 & I3 ($\rho = 0.848$) & 42.81\% & $+0.820$ & $+0.826$ \\
\midrule
Corridor & --- & 39.46\% & $+0.208$ & $+0.000$ \\
\bottomrule
\end{tabular}
\end{table}

\section{Policy Granularity and Schedule Structure}
\label{app:granularity}

Two components of the solver were ablated on the $2 \times 2$ grid of repair and day-type covariate, with $\beta_k$ recalibrated in every cell. Budget repair supplied the larger part of the reported savings: without it the independent MDP reached $29.67\%$ and the cooperative policy $35.64\%$, against $39.46\%$ and $39.67\%$ with it, gains of $9.79$ and $4.03$ percentage points. The day-type covariate was inert at this discretization: the four cells collapse to two values, and adding day type changed neither policy in either repair condition, so it is retained for completeness rather than for any effect it produces.

Table~\ref{tab:slept_share} reports the share of vehicle arrivals each algorithm placed in slept hours, against the SLA budget of $14.29\%$ discussed in Section~\ref{sec:results_energy}.

\begin{table}[t]
\centering
\footnotesize
\caption{Share of vehicle arrivals falling in slept hours (\%) per algorithm, against the SLA budget $\varepsilon/\Pviol = 14.29$\%}
\label{tab:slept_share}
\small
\setlength{\tabcolsep}{3pt}
\begin{tabular}{@{}lrrrrr@{}}
\toprule
Algorithm & I1 & I2 & I3 & I4 & Corridor \\
\midrule
Periodic & 11.91 & 11.93 & 12.57 & 13.15 & 12.35 \\
Threshold & 13.43 & 14.16 & 13.23 & 13.76 & 13.67 \\
MDP & 14.05 & 13.90 & 13.93 & 13.90 & 13.94 \\
Cooperative & 14.05 & 13.83 & 13.77 & 13.84 & 13.87 \\
\bottomrule
\end{tabular}
\end{table}

\begin{figure}[t]
\centering
\includegraphics[width=0.75\columnwidth]{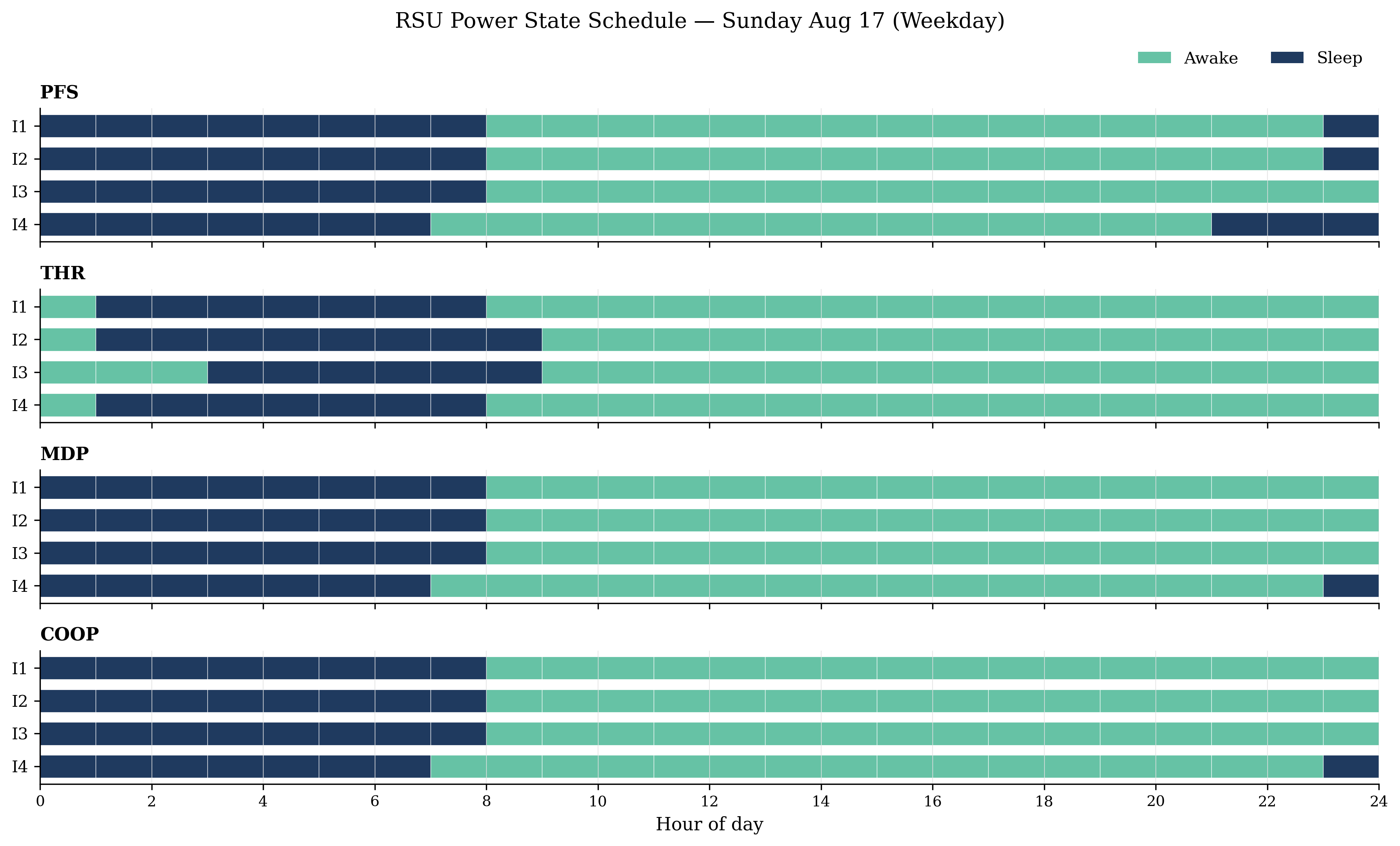}
\footnotesize
\caption{RSU power state schedule for Sunday, August~17 (weekday), showing the two power states Awake and Sleep. PFS sleeps during a fixed overnight block. THR fragments sleep across low-traffic periods. MDP and COOP concentrate sleep in the overnight trough and extend it into the shoulder hours that the remaining violation budget allows.}
\label{fig:gantt_weekday}
\end{figure}

Table~\ref{tab:state_hours} gives the awake/sleep split each algorithm realized per RSU, and Fig.~\ref{fig:gantt_weekday} shows where those hours fall.

\begin{table}[t]
\centering
\footnotesize
\caption{Awake/sleep hours per RSU over the 5-day trace, expanding the budget shares of Table~\ref{tab:slept_share}}
\label{tab:state_hours}
\small
\setlength{\tabcolsep}{3pt}
\begin{tabular}{@{}lrrrr@{}}
\toprule
Algorithm & I1 & I2 & I3 & I4 \\
\midrule
Always-On & 120/0 & 120/0 & 120/0 & 120/0 \\
Periodic & 75/45 & 75/45 & 78/42 & 72/48 \\
Threshold & 74/46 & 73/47 & 81/39 & 72/48 \\
MDP & 71/49 & 71/49 & 75/45 & 67/53 \\
Cooperative & 71/49 & 71/49 & 76/44 & 66/54 \\
\bottomrule
\end{tabular}
\end{table}

\begin{figure}[!t]
\centering
\includegraphics[width=0.95\columnwidth]{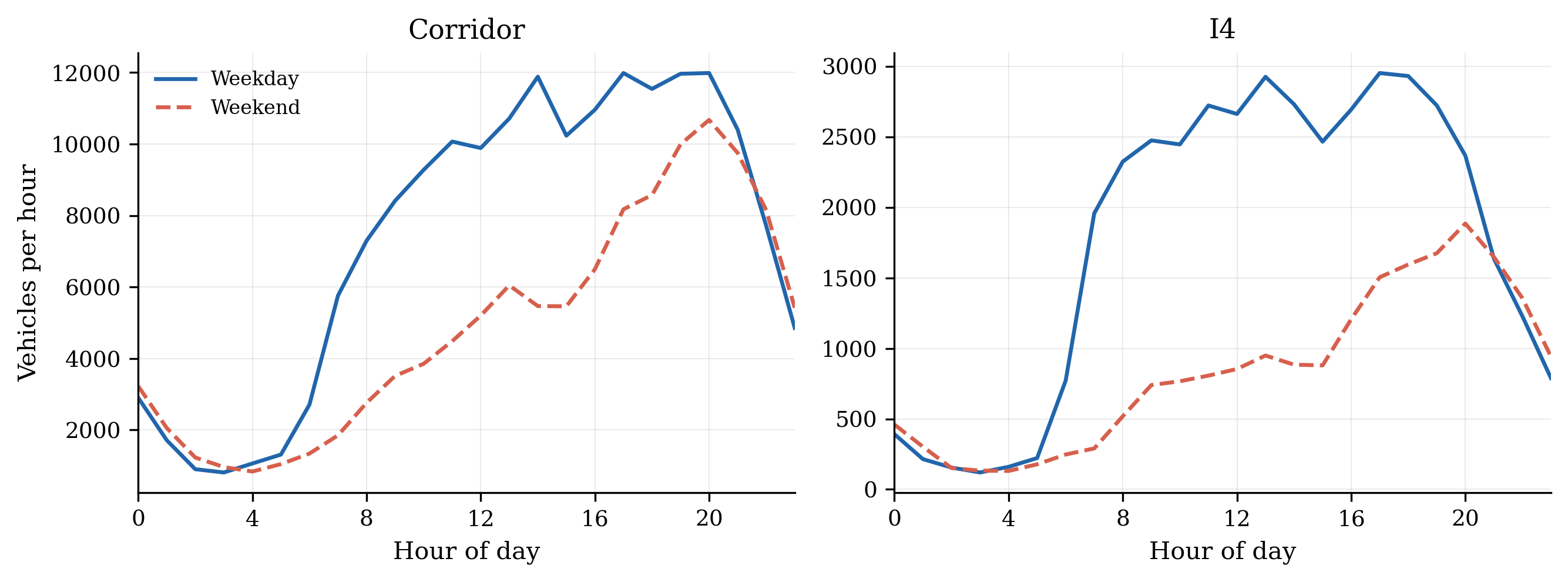}
\footnotesize
\caption{Average hourly traffic profiles, weekday (solid) against weekend (dashed), for the corridor aggregate and for I4, the most volatile intersection. Trough hours (02:00--05:00) offer the primary sleep scheduling window, and the Gulf weekend (Friday--Saturday) shows lower volumes and a later peak. The profiles at I1--I3 follow the corridor shape; their statistics are given in Table~\ref{tab:traffic_stats}.}
\label{fig:traffic_profiles}
\end{figure}

Fig.~\ref{fig:traffic_profiles} plots the hourly traffic profiles whose statistics are tabulated in Section~\ref{sec:traffic_model}.

\begin{figure}[t]
\centering
\includegraphics[width=0.8\columnwidth]{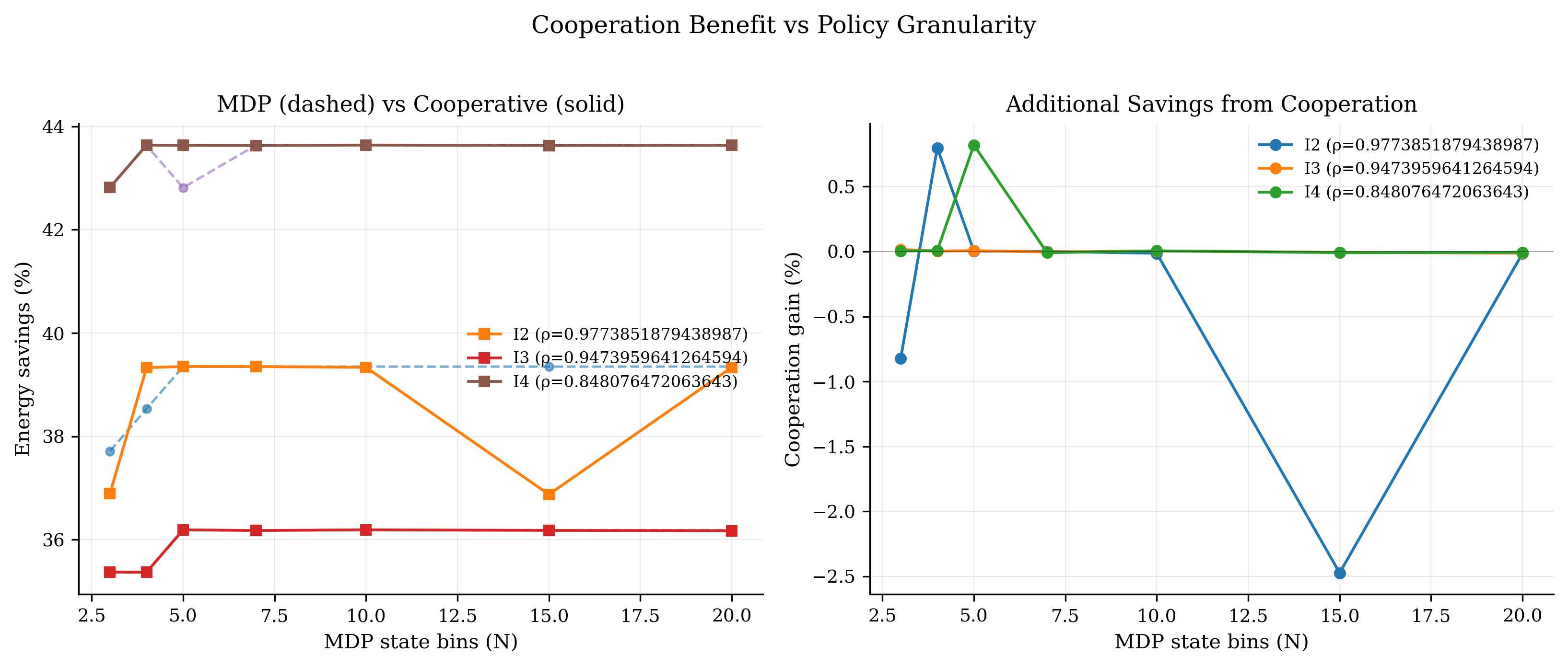}
\footnotesize
\caption{Effect of MDP state granularity, with budget-repair enabled. Left: independent MDP (dashed) and cooperative (solid) energy savings against the number of traffic bins $B$. Right: the difference between them. Both curves are nearly flat for $B \geq 4$, and the cooperative curve does not separate from the independent one at any granularity.}
\label{fig:granularity_sweep}
\end{figure}

The number of traffic bins $B$ sets the resolution of the MDP state and, before the budget-repair was introduced, largely determined the reported savings. Fig.~\ref{fig:granularity_sweep} plots savings against $B$ for the independent and cooperative policies with repair enabled. Both curves are nearly flat for $B \geq 4$, and the cooperative curve does not separate from the independent one at any granularity.

Fig.~\ref{fig:gantt_weekday} shows the corresponding schedules for a representative weekday, with the two power states distinguished by color.

\section{Additional Sensitivity Sweeps and Carbon Accounting}
\label{app:sweeps}

Table~\ref{tab:carbon} expands the carbon figures of Section~\ref{sec:results_carbon} into energy and emissions units for each algorithm.

\begin{table}[t]
\centering
\footnotesize
\caption{Annual carbon footprint reduction against the always-on baseline, at 0.55~kg~CO$_2$/kWh}
\label{tab:carbon}
\small
\setlength{\tabcolsep}{3pt}
\begin{tabular}{@{}lrrrr@{}}
\toprule
& \multicolumn{2}{c}{Per RSU (avg)} & \multicolumn{2}{c}{200-RSU city} \\
\cmidrule(lr){2-3} \cmidrule(lr){4-5}
Algorithm & kWh/yr & kg CO$_2$/yr & kWh/yr & t CO$_2$/yr \\
\midrule
Periodic & 26.7 & 14.7 & 5,336 & 2.93 \\
Threshold & 26.6 & 14.7 & 5,328 & 2.93 \\
MDP & 29.0 & 16.0 & 5,806 & 3.19 \\
Cooperative & 29.2 & 16.1 & 5,838 & 3.21 \\
\bottomrule
\end{tabular}
\end{table}

\begin{figure}[t]
\centering
\includegraphics[width=0.6\columnwidth]{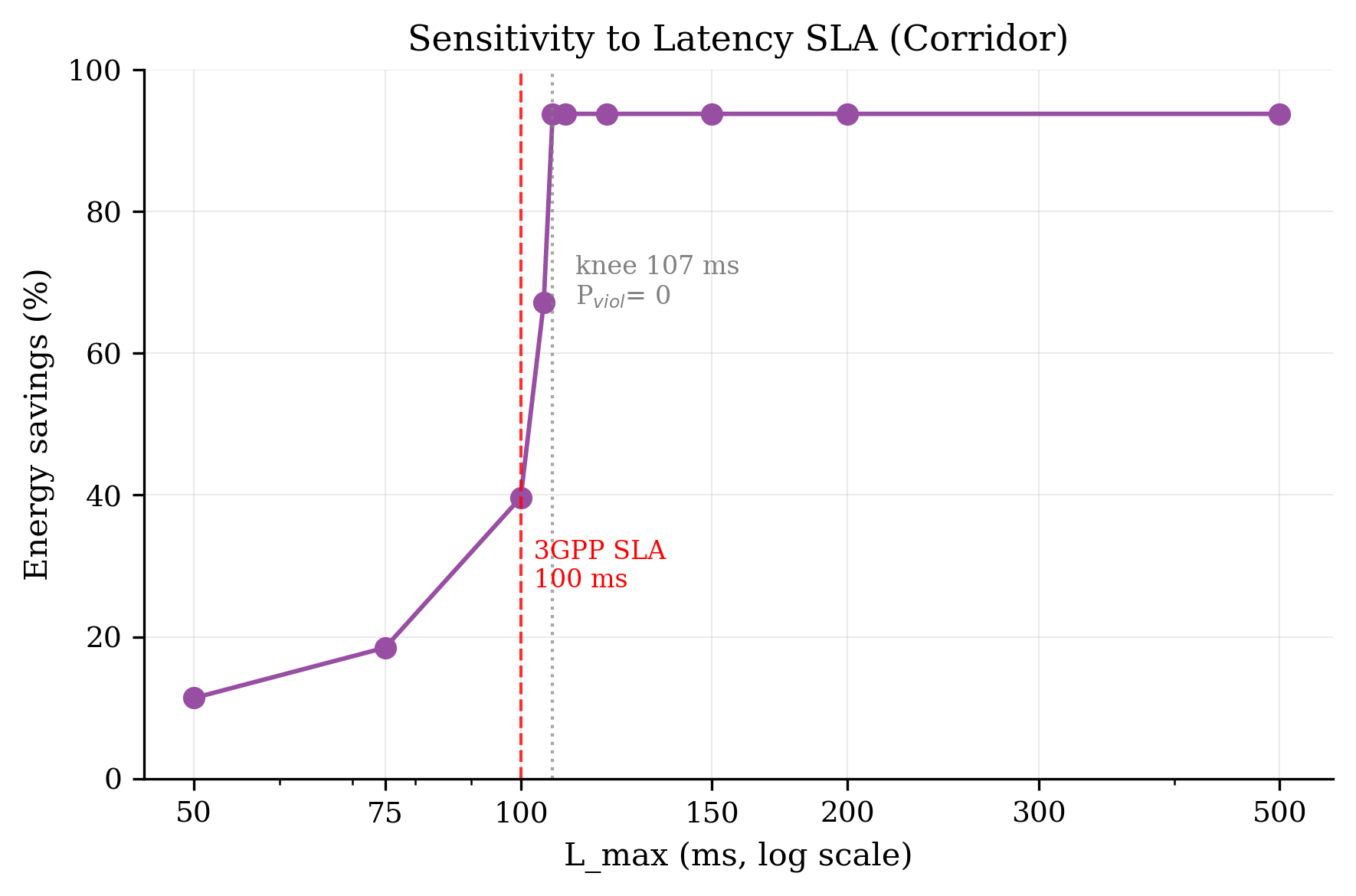}
\footnotesize
\caption{Energy savings vs. latency SLA threshold $\Lmax$. The knee at $\Lmax = \twake + \Ltx + T_{\mathrm{WSA}} = 107$~ms separates the constrained regime ($\Lmax < 107$~ms, where sleep scheduling is a non-trivial optimization) from the unconstrained regime ($\Lmax \geq 107$~ms), where every hour may be slept and savings plateau at $93.77\%$, bounded by the duty-cycle restoration and service energy.}
\label{fig:sensitivity_sla}
\end{figure}

Fig.~\ref{fig:sensitivity_sla} plots the latency-SLA sweep whose knee at $\Lmax = 107$~ms and $93.77\%$ plateau are reported in Section~\ref{sec:results_sensitivity}.

Three of the four parameter sweeps are reported here; the latency SLA sweep, which locates the design boundary, remains in Section~\ref{sec:results_sensitivity}. In each case the policy is fully recalibrated at every point.

\subsection{Wake-Up Delay}

Energy savings degrade gracefully as the hardware wake-up delay increases: from $42.23\%$ at $\twake = 1$~ms to $39.67\%$ at the assumed 2~ms, $34.10\%$ at 5~ms, $27.27\%$ at 10~ms, $20.83\%$ at 20~ms, $11.71\%$ at 50~ms, and $6.42\%$ at 100~ms. The degradation is monotone and has no cliff: at every point in the sweep the portfolio solver still returns a feasible schedule that spends part of the violation budget, with corridor violation rates held between $0.93\%$ and $0.98\%$ throughout. A longer wake-up delay raises $\Pviol$ and so shrinks the vehicle budget, which reduces how many hours can be slept, but it never renders sleep infeasible outright within the swept range. The assumed $\twake = 2$~ms (Section~\ref{sec:power_model}) is therefore not a knife-edge choice, though the sensitivity is real: savings remain above $34\%$ for any $\twake \leq 5$~ms.

\subsection{Traffic Volume Scaling}
\label{sec:sensitivity_traffic}

Retrained policies hold nearly constant savings across the range, from $40.15\%$ at 0.25$\times$ to $38.46\%$ at 3.0$\times$; the budget of Remark~\ref{lem:budget} is a \emph{share} of arrivals, so uniform scaling leaves the feasible schedule set almost unchanged. The fixed 1.0$\times$ policy is reported for robustness, not as an operating point: at 0.25$\times$ it appears to reach $97.05\%$ but at a $7.00\%$ violation rate, because the 1.0$\times$ bin edges place every scaled hour in the lowest bin, and at 3.0$\times$ it under-sleeps to $21.03\%$. The discretization, not the policy structure, is what must be re-fitted when volume shifts, and at $54.4$~s per corridor retraining is cheap enough to run daily.

\subsection{RSU Power Class}

Percentage savings are almost flat across the RSU power class, from $39.67\%$ at $\Pawake = 8.4$~W to $40.15\%$ at 25~W, because both the awake hours and the serving fraction of the slept hours scale with $\Pawake$ and only the residual sleep-power term shifts the ratio. Absolute savings do scale, from 1,599~Wh to 4,817~Wh over the 5-day corridor trace, so higher-power units gain proportionally more from the same schedule.

\label{pg:append}

%
%
%
%